\documentclass{aa}
\usepackage{graphicx}
\usepackage{txfonts}
\usepackage{natbib}
\usepackage{longtable}

\begin{document}

\title{Tracing the merger-driven evolution of active galaxies using the CJF sample}

   \author{M. Karouzos\inst{1}
        \fnmsep\thanks{Member of the International Max Planck Research School (IMPRS) for Astronomy and Astrophysics at the Universities of Bonn and Cologne}
      \and S. Britzen\inst{1}
           \and A. Eckart\inst{2,1}
           \and A. Witzel\inst{1}
           \and A. Zensus\inst{1,2} }

   \institute{ Max-Planck-Institut f\"ur Radioastronomie, Auf dem H\"ugel 69, 53121 Bonn, Germany \\
            \email{mkarouzos@mpifr-bonn.mpg.de}
   \and
   I.Physikalisches Institut, Universit\"at zu K\"oln, Z\"ulpicher Str. 77, 50937 K\"oln, Germany
            }

\offprints{Marios Karouzos}

   \date{Received / Accepted}

\abstract{For the evolution of large-scale structures in the Universe, it is unclear whether active galaxies represent a phase that each galaxy undergoes, and whether and to which extent the evolution of black holes at their centers is important. Binary black hole (BBH) systems may play a key role in our understanding of the above questions.}
{We investigate the Caltech-Jodrell Bank flat-spectrum (CJF) sample to identify evidence that supports the merger-driven evolution scheme of active galaxies, and search for tracer-systems of active galactic nuclei (AGN) evolution and possible BBH candidates. We discuss the strength of and uncertainty in the evidence and formulate a set of selection criteria to detect such tracer-systems.}
{We conduct an extensive literature search for all available  multiwavelength data, particularly in the optical and infrared regime, and morphological information about the CJF sources. We perform a statistical analysis of the properties of this sample.}
{We find 1 ULIRG (Mrk 231) included in the CJF prototype of a transitory system. In total, 28.6\% of the CJF sources with $z\le0.4$ are distorted or have a companion. Given the unbiased sample used here, this provides strong evidence of the ubiquity of the merger phenomenon in the context of active galaxies. We find a correlation between the radio and the near-infrared luminosity of the high-luminosity sources, interpreted in the context of the interplay between a star-formation and AGN component. We find a connection between variability and evolutionary transitory systems, as selected on the basis of their near-infrared colors. We select 28 sources that trace the different evolution phases of an AGN, as well as a number of the most promising BBH candidates. We find 4 sources with almost periodical variability in the optical and radio on similar timescales.}
{}

 \keywords{Galaxies: statistics - Galaxies: active - Galaxies: nuclei - Galaxies: evolution - Galaxies: fundamental parameters }

\authorrunning{Karouzos et al.}
   \maketitle

\section{Introduction}
\label{sec:intro}

As one of the most prominent predictions of the theory of General Relativity, black holes have been at the center of astronomical research for many years. There is a large amount of indirect evidence that most, if not all, galaxies host a supermassive black hole (SMBH) in their center (e.g., \citealt{Blandford1986}).  It is widely believed that this
SMBH is the main constituent of activity in active galactic nuclei (AGN) (for a review see e.g., \citealt{Begelman1984}). Although the mechanism that triggers this activity remains unknown, it has been argued by many authors (e.g., \citealt{Hutchings1983}; \citealt{Sanders1988}; \citealt{Hopkins2006}) that galactic mergers should play a predominant role in these processes.

In the context of the hierarchical model (e.g., \citealt{Frenk1988}), galaxies with a binary black hole (BBH) at their center should be common systems (e.g., \citealt{Merritt2002}; \citealt{Haehnelt2002}).  These systems, although predicted in theory (\citealt{Begelman1980}; for a review on the evolution of BBH see \citealt{Merritt2005}) have been difficult to study. Only a few of them have been directly observed (e.g., \citealt{Hudson2006}; \citealt{Komossa2003}; \citealt{Rodriguez2006}), while in a number of galaxies the presence of a BBH has been discussed (for an overview, see \citealt{Komossa2006}). Active galaxies hosting BBH will be primary candidates for the future gravitational wave space interferometer L.I.S.A. (e.g., \citealt{Sesana2005}).

It is important to identify additional BBH candidates. In this paper, we use the Caltech Jodrell-Bank Flat-spectrum (CJF) flux-limited sample of 293 radio-loud active galaxies to search for these systems, while looking for evidence of the merger-driven evolution scheme for AGN and identifying AGN at different evolutionary stages. This is achieved by means of a multiwavelength analysis of the sample.

This paper is organized as follows. In Sect. \ref{sec:cjf}, we describe the CJF sample. In Sect. \ref{sec:bbh}, we describe the different states of BBH systems, their properties, and the criteria to search for them in a large sample of sources. In Sect. \ref{sec:criteria}, we present the criteria, tracers, with which we investigate the relevance of mergers in AGN. In Sect. \ref{sec:data}, we present multiwavelength data for the CJF sample. In Sect. \ref{sec:results}, we present the analysis of the data presented in Sect. \ref{sec:data}. In Sect. \ref{sec:tracing}, we present the candidates for different evolutionary stages and BBH systems. We conclude with a discussion in Sect. \ref{sec:discussion}, a summary, and an outlook. Throughout the paper, we assume the cosmological parameters $H_{0}=71$ $km s^{-1} Mpc ^{-1}$, $\Omega_{M}=0.27$, and $\Omega_{\Lambda}=0.73$.

\section{The CJF Sample}
\label{sec:cjf}

The CJF sample (\citealt{Taylor1996}) consists of 293 radio-loud active galaxies selected (see Table \ref{tab:cjfproperties}) from three different samples (for details, see \citealt{Britzen2007a}). The sources span a large redshift range (see Fig. 1 in \citealt{Britzen2008}), the farthest object being at a redshift $z=3.889$ (1745+624; \citealt{Hook1995}) and the closest at $z=0.0108$ (1146+596; \citealt{deVaucouleurs1991}). The average redshift of the sample is $z_{avg}=1.254$, $z_{BL Lac,avg}=0.546$, $z_{RG,avg}=0.554$, and $z_{QSO,avg}=1.489$ for BL Lacs, radio galaxies, and QSOs, respectively. All the objects have been observed  with the VLBA and/or the global VLBI network. Each source has at least 3 epochs of observations and has been imaged and studied kinematically (\citealt{Britzen1999}; \citealt{Britzen2007b}; \citealt{Britzen2008}). The X-ray properties have been studied and correlated with their VLBI properties (\citealt{Britzen2007a}).
\begin{table}
\caption{CJF sample and its properties.}
\label{tab:cjfproperties}
\begin{tabular}{l c}
\hline\hline
\textbf{Frequency(MHz)}      &  4850 \\
\textbf{Spectral Index}      &  $\alpha_{1400}^{4850}\geq -0.5$ \\
\textbf{Declination}         &  $\delta\geq 35^{\circ}$  \\
\textbf{Galactic latitude}   &  $|b|\geq 10^{\circ}$ \\
\textbf{\# Quasars}           &  198 \\
\textbf{\# BL Lac }           &  32 \\
\textbf{\# Radio Galaxies}    &  52 \\
\textbf{\# Unclassified}      &  11  \\
\textbf{\# Total}             &  293 \\
\hline
\end{tabular}
\end{table}
\section{Binary black hole system properties and identification criteria}
\label{sec:bbh}

Binary black hole systems are expected to represent an intermediate evolutionary stage of, possibly all (\citealt{Britzen2001}), active galaxies, that are the result of a prior merger (e.g., \citealt{Merritt2005}). The evolutionary track of bound BBH systems has been studied extensively (for a review, see \citealt{Merritt2005}) and is divided in three major phases:
\begin{itemize}
        \item as the two host galaxies merge, the two black holes sink towards the center of the merging system through dynamical friction;
        \item the two black holes form a bound binary system and continue to lose energy by means of gravitational interaction with the central stellar population and its surrounding gas;
        \item and finally, after reaching a critical separation, the systems rapidly lose energy by means of gravitational radiation, leading to the coalescence of the two black holes.
      \end{itemize}
As bound (or "hard") BBH systems, we consider a system of two black holes for which its binding energy per unit mass exceeds the kinetic energy of the stellar population in the nucleus of the system. From the above, we expect the BBH properties to be directly linked to the properties of the host galaxy and the activity of the resulting AGN.

Given the distance of active galaxies and the timescales involved, one expects to observe BBH systems in the first two phases of their evolution,  before or just after they have formed a bound pair. We are therefore interested in identifying and classifying systems that have recently or are currently undergoing a merger phase. According to simulations (e.g.,  \citealt{Mayer2007}), these systems exhibit complex morphologies because of the tidal forces exerted on the merging galaxies.

Ultra luminous infrared galaxies (ULIRGs) probably host an early, unbound BBH system because they are most likely to be merging systems and the precursors of AGN (\citealt{Canalizo2001}), (as is the case of NGC 6240 and 3C 75, \citealt{Komossa2003}; \citealt{Hudson2006}). Almost half of the ULIRGs observed so far exhibit nuclear emission lines, typical of LINER and Seyfert nuclei (e.g., \citealt{Nagar2003}). All of the above phenomena can be linked together (\citealt{Hopkins2006}), illustrating the connection between galaxy, and AGN in particular, evolution and BBH research.

Mergers are not the only systems in which searches for BBH systems have been performed. Binary black holes in the second stage of their evolution may reside in the nuclei of relaxed active galaxies, resembling to first order a single black hole AGN. Very Long Baseline Interferometry (VLBI) techniques have been used to map sources with complicated structures that have been explained in terms of a BBH (e.g., 0420-014, \citealt{Britzen2001}; 3C345, \citealt{Lobanov2005}; 1803+784, \citealt{Roland2008}). A BBH system may produce a precessing jet base (e.g., \citealt{Yokosawa1985}; \citealt{Romero2003}; \citealt{Roland2008}; \citealt{Kaastra1992}) and thus a bent jet. The detected periodicities in the light curves may also be caused by a BBH (e.g., \citealt{Lehto1996}; \citealt{Sillanpaa1988}; \citealt{Qian2007}; \citealt{Caproni2004}). Finally, there is a number of more indirect evidence possibly connected to these systems (e.g., \citealt{Komossa2006}).

\section{Evolution criteria}
\label{sec:criteria}

We search for evidence supporting the connection between activity in galaxies and merger events. We also identify and classify candidate systems at different evolutionary stages using the following criteria:
\begin{itemize}
\item merger signs (morphological distortions, tidal tails, companions etc.);
\item excess infrared fluxes and starburst activity;
\item color information.
\end{itemize}

In the context of the merger-driven evolution of AGN, we search in addition for BBH candidates . We assume that BBH systems are created in systems that have undergone a merger. It is expected that unrelaxed (morphologically disturbed) systems host still unbound BBH systems, whereas bound pairs of black holes are expected to be found in more evolved systems ($\sim10^{6}-10^{7}$ years after the merger, \citealt{Begelman1980}). We use variability as a tracer of these evolved systems. Although the above do not provide robust evidence of a BBH system, they may be used as strong filters for selecting possible candidates.

\subsection{Host galaxy signatures}
\underline{Merger signs} -
We search for morphological information that would characterize a system that is about to, is currently, or has recently undergone a merger event. We use this information to illustrate the connection between activity in galaxies and merger events. Systems with close companions are the precursors of merging systems and represent the earliest evolutionary stages of these systems. In this context, we collect in addition clustering information about our sources, as sources in clusters have a higher probability of taking part in galaxy interactions and mergers. Tidal tails and other morphological distortions are used as explicit signs of current or recent merger activity.\\
\\
\underline{Star-formation activity} -
A higher star-formation rate may be caused by merger events (e.g., \citealt{Barnes1991}). It is known that ULIRGs (\citealt{Sanders1988}) are considered to be a direct result of merger events (e.g., \citealt{Sanders1996}) and are systems that are powered by a strong starburst component (e.g., \citealt{Genzel1998}), although higher luminosity ULIRGs appear to be dominated by a nuclear component. We use starburst activity, as well as extreme infrared luminosities induced by intensive star formation, as additional evidence for systems possibly undergoing a merger.\\
\\
\underline{Color information} -
We use the available near-infrared (HJK photometry) information to identify systems that are possibly associated with merger events. Owing to their high concentration of dust, merging systems appear redder in color than their non-merger counterparts or simple evolved stellar populations. We use this effect to identify evolutionary transitory systems.\\
\\
\underline{Infrared emission} -
Infrared emission from active galaxies originates partly in the host galaxy of the AGN, as a result of thermal emission processes (thermal radiation from heated dust grains and gas). Part of the infrared emission also stems from non-thermal processes, namely synchrotron emission, produced by the same seed population of electrons that produces the radio emission of the compact core and the jets of AGN. Infrared emission can be, partially, used as a tracer of the thermal emission of the AGN host galaxy. High resolution infrared observations can potentially resolve the AGN host and reveal any peculiar morphological features, signs of tidal interactions.\\
\\
\underline{Optical emission} -
The optical and UV continuum observed in AGN originate in the inner part of the galaxy, mostly from the accretion disk and its photosphere, and partly from the jet. The emission is predominantly thermal (as implied by the observation of the big blue bump, \citealt{Krolik1999}) and, based on certain assumptions, the radiation resembles that of a black body, affected however by different absorption mechanisms. The non-thermal jet component may however dominate the total emission because of beaming effects. As in the infrared, any morphological disruptions may also be seen in the optical.

\subsection{AGN signatures}
\underline{Variability} -
Although variability can be induced by a number of different effects, almost periodical, multiwavelength variations in the lightcurves of objects point towards a periodic modulation in the nucleus of the system. Assuming that a hard BBH system is possibly the origin of these variations, we use multiwavelength variability as a tracer of hard BBH systems and thus late-merger (or merger-remnant) systems.\\
\\
\underline{Radio Emission} -
Radio emission originates mostly in the compact core and jet of the active galaxy. It is produced by relativistic electrons, gyrating in a magnetic field, by means of synchrotron radiation. Radio properties of a source provide direct information about the properties of the AGN. The radio luminosity of an active galaxy can be used as a measure for its non-thermal, AGN component.\\
\\
A more detailed discussion of the above can be found later on in Sect. \ref{sec:discussion}.

\section{ Multiwavelength properties of the CJF sample}
\label{sec:data}

We present multiwavelength information about the CJF sources gathered from our extensive search of the available literature. We focus on the radio, infrared, and optical parts of the spectrum.

\subsection{Radio emission}

The CJF is a flux-limited radio-selected sample of flat-spectrum radio-loud AGN. Although originally created to study the kinematics of pc-scale jets and superluminal motion, the CJF represents an ideal sample for an unbiased study of the evolution of active galaxies. Unlike previously used samples in the same context, the radio-selection of the CJF allows us to exclude biases towards transitory systems (as in the case of infrared-selected samples).

The CJF sample (Table \ref{tab:cjfproperties}) has been most extensively studied in the radio regime (e.g., \citealt{Taylor1996}; \citealt{Pearson1998}; \citealt{Britzen1999}; \citealt{Vermeulen2003}; \citealt{Pollack2003}; \citealt{Lowe2007}; \citealt{Britzen2007a}; \citealt{Britzen2007b}; \citealt{Britzen2008}). \citet{Britzen2008} develop a localized method for calculating the bending of the jet associated with individual components. The maximum of the distribution of local angles is at zero degrees, although a substantial fraction shows some bending ($0-40$ degrees). A few sources exhibit sharp bends of the order of $>50$ degrees (see Fig. 13 in \citealt{Britzen2008}). These morphologies on pc-scales (for a detailed discussion, see \citealt{Britzen2008}) can be explained in the context of a BBH system. A future paper will study the morphologies of the jets of the CJF sources and their evolution.

Although the CJF sample consists mostly of blazars, presumably highly beamed sources, the kinematical study of the sample identifies a large number of sources with stationary, subluminal, or, at best, mildly superluminal outward velocities (e.g., see Fig. 15 in \citealt{Britzen2008}). Combined with a number of sources with inwardly moving components (e.g., 0600+422, 1751+441, 1543+517, \citealt{Britzen2007b}), these sources do not fit into the regular paradigm of outward, superluminaly moving components in blazar jets. One explanation of these peculiar kinematic behaviors is that of a precessing, or helical, jet (e.g., \citealt{Conway1993}) as a result of a BBH system. Other interpretations include trailing shocks produced in the wake of a single perturbation propagating down the jet (e.g., \citealt{Agudo2001}; but also see \citealt{Mimica2009}), or standing re-collimation shocks (e.g., \citealt{Gomez1995}).

Using the power spectral analysis method, \citet{Fan2007} study the lightcurves of a large number of AGN at three different frequencies, showing that there are different periodicities in them. There are 19 CJF sources in the Fan et al. sample (see Table \ref{tab:multivar}) that have almost periodical behavior in their radio lightcurves. Seven additional sources exhibit almost periodical variability in their radio lightcurves, as found in the literature (in total 26 sources, 8.9\% of the sample; Tables \ref{tab:multivar} and \ref{tab:multivarref} for references). Intra-day variability is not considered here (see Sect. \ref{sec:discussion} for details).

\begin{table*}
\begin{center}
\caption{CJF sources studied for variability across the spectrum. If available, we indicate the respective timescales. Columns (1)-(4) give the IAU source name, other name, the type of the source, and its redshift, respectively, Col. (5) gives the variability in the radio, Col. (6) in the infrared, Col. (7)in the optical, Col. (8) in the X-ray, and Col. (9) in the $\gamma$-rays. For references, see Table \ref{tab:multivarref}.}
\label{tab:multivar}
\begin{tabular}{l l l l l l l l l}
\hline\hline
\textbf{Source}	&	\textbf{Other Names}	&\textbf{Type}	&\textbf{z}	&\textbf{Var}$\mathbf{_{Radio}^{1}}$	 &\textbf{Var}$\mathbf{_{IR}}$	 &\textbf{Var}$\mathbf{_{Opt}}$	&\textbf{Var}$\mathbf{_{Xray}}$	&\textbf{Var}$\mathbf{_{\gamma}}$ \\
        &               &       &   &       (yr)        &       (yr)            &       (yr)            &       (days)         &       (days)               \\
        \hline
 0016+731&			&Q	  &1.781		&6.6			           &	                  &                                                     &             &             \\
 0035+367&	4C 36.01&Q	  &0.366		&        			       &			          &6.8$^{18}$                                           &             &             \\
 0133+476&	OC 457	&Q    &0.859		&5.8-10.9		           &		              &                                                     &             &             \\
 0212+735&			&Q    &2.367		&15			               &	                  &                                                     &             &             \\
 0219+428&	3C 66A	&BL   &0.444		&                          &yes$^{9}$            &4.25$^{10}$, 2.5$^{19}$, 63$^{20,**}$, 65$^{21,**}$	&             & yes$^{39}$  \\
 0316+413&	3C 84	&G	  &0.018		&9.7			           &	                  & 3.06, 65-68$^{22}$   	                            &800$^{34,*}$ &             \\
 0402+379&	4C 37.11&G    &0.055		&yes$^{2}$        		   &	                  &                                                     &             &             \\
 0454+844&			&BL	  &0.112		&			               &			          &3.1$^{232,**}$	                                    &		      &		        \\
 0642+449&	OH+471	&Q    &3.396		&			               &			          &yes$^{24}$		                                    &yes$^{35}$	  &             \\
 0711+356&	OI 318	&Q    &1.620		&0.7$^{3}$		           &	                  &11.2$^{25}$                                          &             &             \\
 0716+714&			&BL   &		        &5.4-15,5.5-6$^{4}$       &yes$^{10}$            &3.3$^{4}$                                           &1470$^{34,*}$&yes$^{40}$   \\
 0804+499&	OJ508	&Q    &1.432        &1.1, 1.8, 2.7$^{3}$	   &                      &                                                     &             &yes$^{36}$   \\
 0814+425&	OJ425	&BL   &0.245		&19.2-20.7		           &yes$^{11}$	          &2.4$^{23,**}$                                        &             &             \\
 0836+710& 4C 71.07 &Q    &2.180		&5.7-9.0		           &		              &                                                     &             &yes$^{39}$   \\
 0917+449&VIPS 0214 &Q	  &2.180		&			               &			          &			                                            &		      &yes$^{39}$   \\
 0923+392&	4C 39.25&Q    &0.699		&22.5-22.4		           &(?)$^{12}$	          &	yes$^{22}$		                                    &             &             \\
 0927+352&          &BL   &             &6$^{5}$                  &                      &                                                     &             &             \\
 0945+408&			&Q    &1.252		&3.6-7.3$^{6}$	           &                      &                                                     &             &             \\
 0954+556&	4C 55.17&Q    &0.909		&			               &			          &			                                            &		      &yes$^{39}$   \\
 0954+658&			&BL	  &0.368		&			               &			          &2.6$^{23,**}$	                                    &		      &yes$^{39}$	\\
 1038+528&	OL 564	&Q    &0.677		&9.3			           &	                  &                                                     &             &             \\
 1101+384&	Mrk 421	&BL	  &0.031		&3.7-8.0		           &yes$^{13,14}$         &	6.2$^{18}$	                                        &62$^{36}$    &yes$^{39}$   \\
 1254+571&  Mrk 231 &G    &0.042        &yes$^{7}$                &                      &yes$^{26}$                                           &yes$^{37}$   &             \\
 1418+546&	OQ530	&BL   &0.151		&2.5-11.3	               &yes$^{13}$	          &	yes$^{27}$		                                    &             &             \\
 1504+377&			&G	  &0.671		&			               &yes$^{15}$            &			                                            &		      &			    \\
 1624+416&	4C 41.32&Q    &2.550		&16.7-37.1		           &		              &                                                     &             &             \\
 1633+382&	4C 38.41&Q    &1.807		&14.3			           &			          &	yes$^{22}$		                                    &             & yes$^{39}$  \\
 1638+398&	NRAO 512&Q	  &1.660		&			               &			          &	0.5-31$^{22}$		                                &             &             \\
 1642+690&			&Q    &0.751		&6.6-7.0		           &			          &                                                     &             &             \\
 1641+399&	3C345	&Q    &0.595        &5.1$^{3}$, 3.5-4$^{8}$  &yes$^{13}$            &5.6, 11.4$^{28}$, 10.1$^{29}$, 1.9$^{30}$            &0.72$^{34}$  &             \\
 1652+398&	Mrk 501	&BL   &0.034	    &6.8-12.3	               &yes$^{8}$	          &yes$^{27}$                                           &23$^{38}$    &23.2$^{41}$  \\
 1739+522&	4C 51.37&Q    &1.381		&1.4$^{3}$		           &		              &                                                     &             &yes$^{39}$   \\
 1749+701&			&BL   &0.769		&12.3			           &  			          &	yes$^{22}$		                                    &             &             \\
 1803+784&			&BL   &0.679		&7.0-11.3		           &		              & 1.1$^{23,**}$                                       &             &             \\
 1807+698& 3C 371.0 &BL   &0.051		&			               &yes$^{16}$            &2.7$^{10}$, 2.4$^{18}$	                            &             &             \\
 1823+568&			&BL   &0.664		&8.8-14.9		           &		              &                                                     &             &             \\
 1928+738& 4C 73.18 &Q    &0.302		&7.7-8.2		           &			          &	yes$^{22}$		                                    &	          &             \\
 2007+777&			&BL	  &0.342		&			               &yes$^{17}$	          &0.6$^{23,**}$, 10-40$^{31,**}$                       &             &             \\
 2200+420&	BL Lac	&BL   &0.069		&3.9-7.8                   &yes$^{13}$	          &7.8$^{32}$, 0.6, 0.88, 14$^{33}$, 7.7$^{18}$         &             &yes$^{39}$   \\
 2351+456& 4C 45.51 &Q    &1.986		&11.8			           &	                  &                                                     &             &             \\
\hline
\end{tabular}
\end{center}
\end{table*}

\begin{table*}
\begin{center}
\caption{References for Table \ref{tab:multivar}.}
\label{tab:multivarref}
\begin{tabular}{l l l l}
\hline\hline
\textbf{Symbol}   &    \textbf{Reference}              & \textbf{Symbol} &     \textbf{ Reference }              \\
\hline
1       &    \citet{Fan2007}        &22     &    \citet{Bozyan1990}        \\
2       &    \citet{Rodriguez2006}  &23     &   \citet{Heidt1996}          \\
3       &    \citet{Kelly2003}      &24     &   \citet{Pica1983}           \\
4       &    \citet{Raiteri2003}    &25     &   \citet{Smith1993}          \\
5       &    \citet{Machalski1996}  &26     &   \citet{Kollatschny1992}    \\
6       &    \citet{Ciaramella2004} &27     &   \citet{Iwasawa1998}        \\
7       &    \citet{Ulvestad1999b}  &28     &   \citet{Webb1988}           \\
8       &    \citet{Lobanov1999}    &29     &   \citet{Zhang1998}          \\
9       &    \citet{Abdo2009}       &30     &   \citet{Babadzhanyants1995} \\
10      &    \citet{HagenThorn2006} &31     &   \citet{Bychkova2003}       \\
11      &    \citet{Fan1999}        &32     &   \citet{HagenThorn1997}     \\
12      &    \citet{Neugebauer1986} &33     &   \citet{Fan1998}            \\
13      &    \citet{Takalo1992}     &34     &   \citet{BianZhao2003}       \\
14      &    \citet{Gupta2004}      &35     &   \citet{Zamorani1984}       \\
15      &    \citet{Stickel1996b}   &36     &   \citet{Osone2001}          \\
16      &    \citet{Edelson1987}    &37     &   \citet{Gallagher2002}      \\
17      &    \citet{Peng2000}       &38     &   \citet{Osone2006}          \\
18      &    \citet{Smith1995}      &39     &   \citet{Fan2002}            \\
19      &    \citet{Belokon2003}    &40     &   \citet{Chen2008}           \\
20      &    \citet{Xie2002}        &41     &   \citet{Neshpor2000b}       \\
21      &    \citet{Lainela1999}    &**     &   in days                    \\
\hline
\end{tabular}
\end{center}
\end{table*}

\subsection{Infrared emission}
There has been ongoing discussion about the evolutionary connection between ULIRGs and AGN, with ULIRGs being assumed to be the predecessors of active galaxies by means of a merging process (e.g., \citealt{Canalizo2001}; \citealt{Bekki2006}; \citealt{Riechers2008}; \citealt{Bennert2008}; \citealt{Hopkins2006}). We find that one source in our sample can be classified as an ULIRG (Mrk 231, \citealt{Cao2008}; \citealt{Kim2002}). There are 101 sources (34.5\% of the sample; Table \ref{tab:infrared}) that have been observed and detected (we note that the presented data here are not statistically complete) in the near to far infrared by the Infrared Astronomical Satellite (IRAS, \citealt{Clegg1980}) and the 2 Micron All Sky Survey (2MASS, \citealt{Kleinmann1992}), as well as other Earth-based telescopes.

\begin{table*}[htp]
\begin{center}
\caption{CJF sources observed and detected in the near infrared. Column (1) gives the IAU name of the source, Col. (2) the type of source (Q=Quasar, BL=BL Lac object, G=Radio Galaxy), Col. (3) the redshift z, Cols. (4)-(6) fluxes in three different NIR bands, Col. (7) the relevant references (Table \ref{tab:infraref}), and Col. (8) gives the calculated spectral index in the NIR.}
\label{tab:infrared}
\begin{tabular}{l l l l l l l l}
\hline\hline
\textbf{Source}	&	\textbf{Type}	&\textbf{z}     &    	  	&  \textbf{ \underline{Near-IR}}	&   &\textbf{Reference}    &$\alpha_{NIR}$	\\
        &               &      &      $1.25\mu m$	&   $1.65\mu m$ &   $2.20\mu m$ &     &	\\
	    &		    	&	   &         (mJy)	    &	(mJy)	    &	(mJy)	    &		            &		      	 	\\
\hline
0022+390&	Q	&1.946	&			        &		        &0.06	(0.03)	&42                        &          	   \\
0035+413&	Q	&1.353	&	0.64	(0.06)	&0.75	(0.13)	&1.37	(0.24)	&43                        &   -1.29       \\
0108+388&	G	&0.669	&			        &		        &0.14	(0.06)	&44                        &               \\
0133+476&   Q   &0.859  &   0.479   (0.02)  &0.813  (0.04)  &1.32   (0.04)  &45                        &    -1.74      \\
0145+386&	Q	&1.442	&	0.40	(0.04)	&0.68	(0.11)	&0.30	(0.17)	&43                         &   0.55        \\
0153+744&	Q	&2.388	&	0.93	(0.07)	&1.29	(0.15)	&1.96	(0.23)	&43                         &   -0.89       \\
0219+428&	BL 	&0.444	&	9.22	(0.25)	&12.5	(0.4)	&15.4	(1.1)	&46, 11                      &   -0.83       \\
0249+383&	Q	&1.122	&			        &		        &0.12	(0.03)	&42                         &               \\
0251+393&	Q	&0.289	&	0.70	(0.05)	&0.79	(0.11)	&1.21	(0.19)	&43                         &   -0.89       \\
0316+413&	G	&0.018	&	46.5	(0.7)	&57.8	(0.9)	&55.70	(0.8)	&47			               &	-0.10      \\
0344+405&	G	&0.039	&	17.8	(0.6)	&26.0 	(1.0)   &20.0   (1.0)   &48                         &   0.26        \\
0402+379&	G	&0.055	&	19.1	(1.8)	&39.8	(2.6)	&40 	(3)    	&43                         &   -0.38       \\
0546+726&	Q	&1.555	&	0.84	(0.02)	&1.24	(0.03)	&0.91	(0.02)	&43                         &   0.03        \\
0650+453&	Q	&0.933	&	0.33	(0.02)	&0.49	(0.02)	&0.84	(0.02)	&43                         &   -1.58       \\
0651+410&	G	&0.022	&	46.6	(1.1)	&57.0   (1.9)	&48.0	(1.5)	&48                         &   0.03        \\
0707+476&	Q	&1.292	&	13.2	(0.3)	&10.9	(0.5)	&7.4	(0.6)	&43                         &   1.08        \\
0710+439&	G	&0.518	&			        &		        &0.84	(0.19)	&49                         &               \\
0731+479&	Q	&0.782	&	0.58	(0.02)	&0.52	(0.03)	&0.67	(0.02)	&43                         &   -0.17       \\
0733+597&	G	&0.041	&	43.2	(1.3)	&52.3	(1.9)	&45.1	(1.8)	&48                         &   -0.01       \\
0814+425&	BL 	&0.245	&	1.42	(0.04)	&2.21	(0.06)	&3.14	(0.07)	&11                         &   -1.34       \\
0821+394&	Q	&1.216	&	0.77	(0.02)	&1.15	(0.02)	&1.45	(0.02)	&43                         &   -1.07       \\
0824+355&	Q	&2.249	&			        &		        &0.10	(0.06)	&50                         &               \\
0831+557&	G	&0.240	&	0.80		    &1.03		    &1.23		    &51			               &   -0.72       \\
0833+416&	Q	&1.298	&	0.36	(0.02)	&0.43      		&0.35	(0.02)	&43                         &   0.09        \\
0836+710&	Q	&2.180	&	1.00	(0.02)	&1.30	(0.03)	&1.50	(0.02)	&43                         &   -0.68       \\
0847+379&	G	&0.407	&			        &		        &0.63	(0.04)	&50                         &               \\
0923+392&	Q	&0.699	&	1.22	(0.06)	&1.20	(0.12)	&1.66	(0.22)	&43                         &   -0.53       \\
0927+352&	BL 	&	    &			        &		        &0.81	(0.09)	&50                         &               \\
0945+664&	G	&0.850	&	0.030	(0.001)	&0.005 (0.001)	&0.08   (0.01)	&15                        &   -1.89       \\
0954+658&	BL 	&0.368	&			        &3.17		    &		        &52                        &               \\
1003+830&	G	&0.322	&	0.86	(0.13)	&2.0	(0.4)	&2.3	(0.4)	&48                         &    -1.67      \\
1020+400&	Q	&1.254	&	0.74	(0.04)	&1.00	(0.07)	&1.53	(0.07)	&43                         &    -1.27      \\
1030+415&	Q	&1.120	&	0.07		    &0.12		    &0.30		    &53                        &    -2.56      \\
1031+567&	G	&0.460	&			        &		        &0.19	(0.13)	&49                         &               \\
1038+528&	Q	&0.677	&	0.65	(0.05)	&0.42	(0.10)	&0.84	(0.16)	&43                         &    -0.45      \\
1058+726&	Q	&1.460	&	0.42	(0.02)	&0.44	(0.02)	&0.48	(0.03)	&43                         &    -0.20      \\
1101+384&	BL 	&0.031	&	71	(5)	        &81   	(8)    	&83    	(15)	&11                         &    -0.25      \\
1124+455&	Q	&1.811	&	0.57	(0.01)	&0.54	(0.02)	&0.66	(0.02)	&43                         &    -0.25      \\
1128+385&	Q	&1.733	&			        &		        &0.32		    &53                        &               \\
1144+402&	Q	&1.088	&	1.17	(0.06)	&1.69	(0.15)	&2.53	(0.25)	&43                         &    -1.35      \\
1144+352&	G	&0.063	&	20.0	(0.8)	&34.0	(1.3)	&23.3	(1.5)	&44                         &    -0.24      \\
1146+596&	G	&0.011	&	247	    (2)	    &311	(5)	    &248	(8)	    &43                         &    0.03       \\
1226+373&	Q	&1.515	&	0.33	(0.02)	&0.48	(0.02)	&0.41	(0.03)	&43                         &    -0.36      \\
1254+571&	Q	&0.042  &	62.0	(1.2)	&113	(2)	    &199	(3)	    &54                        &    -2.04      \\
1306+360&	Q	&1.055	&	0.23	(0.02)	&0.41	(0.02)	&0.54	(0.02)	&43                         &    -1.49      \\
1309+555&	Q	&0.926	&	0.65	(0.02)	&0.93	(0.03)	&1.36	(0.03)	&43                         &    -1.28      \\
1333+459&	Q	&2.449	&	0.34	(0.05)	&0.40	(0.11)	&0.66	(0.20)	&43                         &    -1.15      \\
1347+539&	Q	&0.980	&	0.47	(0.02)	&0.53	(0.03)	&0.65	(0.04)	&43                         &    -0.56      \\
1413+373&	Q	&2.360	&	0.36	(0.02)	&0.36	(0.02)	&0.84	(0.03)	&43                         &    -1.50      \\
1418+546&	BL 	&0.151	&	6.68	(0.31)	&9.2	(0.4)	&11.8	(0.5)	&11, 55                    &    -0.99      \\
\hline
\end{tabular}
\end{center}
\end{table*}
\addtocounter{table}{-1}
\begin{table*}
\begin{center}
\caption{Continued.}
\begin{tabular}{l l l l l l l l l l}
\hline\hline
\textbf{Source}	&	\textbf{Type}	&\textbf{z}     &    	  	&  \textbf{ \underline{Near-IR}}	&   	& \textbf{Reference}   &$\alpha_{NIR}$	\\
        &               &      &      $1.25\mu m$	&   $1.65\mu m$ &   $2.20\mu m$ &     &	\\
	    &		    	&	   &         (mJy)	    &	(mJy)	    &	(mJy)	    &		            &		      	 	\\
\hline
1417+385&	Q	&1.832	&	0.41	(0.02)	&0.52	(0.02)	&0.66	(0.02)	&43                         &    -0.83      \\
1432+422&	Q	&1.240	&	0.02		    &		        &0.05		    &56                        &               \\
1435+638&	Q	&2.068	&	1.10	(0.02)	&1.00	(0.03)	&1.40	(0.03)	&43                         &    -0.42      \\
1442+637&	Q	&1.38	&	0.51	(0.02)	&0.88	(0.02)	&0.71	(0.02)	&43                         &    -0.56      \\
1504+377&	G	&0.671	&	0.06	(0.01)	&0.12	(0.01)	&0.25	(0.02)	&15			               &    -2.62      \\
1531+722&	Q	&0.899	&	1.25	(0.12)	&2.0	(0.4)	&1.7	(0.4)	&48                        &    -0.53      \\
1622+665&	G	&0.201	&			        &		        &3.3	(0.3)	&57                        &               \\
1633+382&	Q	&1.807	&			        &		        &1.95	(0.07)	&49, 58                     &               \\
1637+574&	Q	&0.749	&	1.70	(0.02)	&1.90	(0.02)	&2.50	(0.03)	&12                        &    -0.67      \\
1641+399&	Q	&0.595	&	3.23	(0.10)	&5.08	(0.25)	&7.5	(0.4)	&43                         &    -1.46      \\
1652+398&	BL 	&0.034  &	86.0	(2.0)  	&115.0	(2.0)   &99.0   (2.0)   &59, 11                     &    -0.22      \\
1656+477&	Q	&1.622	&	0.53	(0.02)	&0.60	(0.03)	&0.53     		&43                         &    0.02       \\
1700+685&	G	&0.301	&	0.64	(0.02)	&1.00	(0.03)	&1.60	(0.04)	&43                         &    -1.57      \\
1716+686&	Q	&0.339	&	1.30       		&1.70     		&2.00     		&43                         &    -0.73      \\
1719+357&	Q	&0.263	&	0.83	(0.21)	&1.6	(0.3)	&2.8	(0.4)	&48                         &    -2.10      \\
1726+455&	Q	&0.717	&	0.61	(0.02)	&0.84	(0.03)	&1.01	(0.03)	&43                         &    -0.86      \\
1732+389&	Q	&0.970	&	0.26	(0.02)	&0.49	(0.02)	&0.77	(0.02)	&43                         &    -1.87      \\
1744+557&	G	&0.031	&	66     	(1)    	&80    	(2)    	&66    	(2)    	&60                        &    0.05       \\
1749+701&	BL 	&0.770	&			        &		        &2.2	(1.7)	&55                        &               \\
1755+578&	Q	&2.110	&	0.33	(0.02)	&0.50	(0.02)	&0.67	(0.03)	&43                         &    -1.20      \\
1758+388&	Q	&2.092	&	0.43	(0.02)	&0.38	(0.02)	&0.49	(0.02)	&43                         &    -0.21      \\
1803+784&	BL 	&0.680	&	5.50	(0.25)	&5.9	(0.4)	&7.1	(0.8)	&55,11                     &    -0.39      \\
1800+440&	Q	&0.663	&	1.00       		&1.20     		&0.84	(0.02)	&43                         &    0.37       \\
1807+698&	BL 	&0.051	&	20.2	(0.9)	&25.4	(1.1)	&29.7	(2.7)	&59, 11                     &    -0.65      \\
1826+796&	G	&0.224	&	1.10	(0.20)	&1.3	(0.3)	&1.9	(0.4)	&48                         &    -0.89      \\
1823+568&	BL 	&0.664	&			        &		        &4.98   		&11, 61                     &               \\
1843+356&	G	&0.764	&			        &		        &0.13	(0.09)	&44                         &               \\
1946+708&	G	&0.101	&	2.8    	(0.3)	&2.4	(0.5)	&3.0	(0.5)	&57                        &    0.07       \\
2116+818&	Q	&0.084	&	8.5    	(0.4)	&12.5	(0.6)	&15.7	(0.7)	&48                         &    -0.95      \\
2200+420&	BL 	&0.069	&	38.4	(2.8)	&53   	(4)    	&71    	(14)	&43, 11                      &    -0.80      \\
2351+456&	Q	&1.986	&			        &		        &0.63 		    &49                         &               \\
2352+495&	G	&0.237	&			        &		        &0.9    (0.4)	&49                         &               \\
\hline
\end{tabular}
\end{center}
\end{table*}

\begin{table*}[htp]
\begin{center}
\caption{References for Table \ref{tab:infrared}.}
\label{tab:infraref}
\begin{tabular}{l l l l}
\hline\hline
\textbf{Number} &    \textbf{Reference}   &   \textbf{Number}      &\textbf{Reference} \\
\hline
42      &    \citet{Sanchez2003}    &  52      &   \citet{Wright1998}       \\
43      &   \citet{Barkhouse2001}   &  53      &   \citet{Carballo1998}     \\
44      &   \citet{Snellen1996}     &  54      &   \citet{Sanders1988}      \\
45      &   \citet{Bloom1994}       &  55      &   \citet{Padovani2006}     \\
46      &   \citet{Gonzalez2001}    &  56      &   \citet{Puschell1982}     \\
47      &   \citet{Jarrett2003}     &  57      &   \citet{Villani1999}      \\
48      &   \citet{Skrutskie2003}   &  58      &   \citet{Xie1998}          \\
49      &   \citet{deVries1998}     &  59      &   \citet{Golombek1988}     \\
50      &   \citet{Lilly1985}       &  60      &   \citet{Knapp1990}        \\
51      &   \citet{Heckman1983}     &  61      &   \citet{Kotilainen2005}   \\
\hline
\end{tabular}
\end{center}
\end{table*}

Only a few blazars have extensive light curves in the infrared. We identify 12 sources (4.1\% of the sample) with infrared variability (Table \ref{tab:multivar}).

\subsection{Optical emission}

Almost all CJF sources (285 sources, 97.3\% of the sample) have been detected in the optical (see Table 1 in \citealt{Britzen2007a} for magnitudes and redshifts).
\begin{table*}[htp]
\begin{center}
\caption{CJF sources observed and detected in the optical regime as group or cluster members. Cols. (1)-(3) as in Table \ref{tab:infrared}, Col. (4) gives the reference for clustering (Table \ref{tab:opticalref}).}
\label{tab:clusters}
\begin{tabular}{l l l c| l l l c}
\hline\hline
\textbf{Source}	        &	\textbf{Type}    &\textbf{z}	       &\textbf{Reference}   &\textbf{Source}     &\textbf{  Type}	    &  \textbf{z}	   &	 \textbf{Reference}   \\
\hline
0010+405        &	G		&0.255	   &      62    &1038+528	&	Q		&0.677	   &	81          \\
0014+813	    &	Q		&3.366	   &	  63   &1101+384   &	BL 		&0.031	   &    82          \\
0153+744        &	Q		&2.388	   &      64    &1144+352	&	G		&0.063	   &    83          \\
0212+735        &	Q		&2.367	   &      64    &1146+596   &	G		&0.011	   &    84   	    \\
0219+428        &	BL 		&0.444	   &      65    &1150+812   &	Q		&1.250	   &    64          \\
0309+411        &	Q		&0.134	   &      66    &1213+350	&	Q	  	&0.857	   &    62          \\
0316+413	    &	G		&0.018	   &      67    &1216+487	&   Q	  	&1.076	   &    64          \\
0636+680        &	Q		&3.174	   &      68    &1254+571	&	G		&0.422     &    85          \\
0651+410	    &	G		&0.022	   &      69    &1415+463	&   Q	  	&1.552	   &    76          \\
0700+470	    &	G		&	       &      70    &1435+638	&   Q	  	&2.068	   &    65          \\
0710+439        &	G		&0.518	   &      62    &1624+416	&   Q	  	&2.550	   &    77          \\
0727+409        &	Q		&2.500	   &      71    &1633+382	&   Q	  	&1.807	   &    76          \\
0733+597        &	G		&0.041	   &      72    &1636+473	&   Q	  	&0.740	   &    76          \\
0821+394        &	Q		&1.216	   &      73    &1637+574	&   Q	  	&0.749	   &    86          \\
0831+557        &	G		&0.240	   &      74    &1638+540	&	Q       &1.977     &    87          \\
0836+710        &	Q		&2.180	   &      62    &1638+398	&   Q	  	&1.660	   &    79          \\
0847+379	    &	G		&0.407	   &      75    &1642+690	&   Q	  	&0.751	   &    64          \\
0850+581        &	Q		&1.322	   &      73    &1641+399	&   Q	  	&0.595	   &    88          \\
0859+470        &	Q		&1.462	   &      76    &1652+398	&   BL 	  	&0.034     &    89          \\
0923+392        &	Q		&0.699	   &      62    &1716+686	&   Q	  	&0.339	   &    64          \\
0945+408        &	Q		&1.252	   &      65    &1744+557	&	G		&0.031	   &    69          \\
0954+556        &	Q		&0.909	   &      77    &1751+441	&   Q	  	&0.871	   &    64          \\
0955+476        &	Q		&1.873	   &      78    &1803+784	&   BL 	  	&0.680	   &    65          \\
1010+350        &	Q		&1.414	   &      79    &1807+698	&   BL 	  	&0.051	   &    82          \\
1020+400        &	Q		&1.254	   &      76    &2214+350	&   Q	  	&0.510	   &    90          \\
1031+567        &	G		&0.460	   &      80    &          	&   	  	&	       &                \\
\hline
\end{tabular}
\end{center}
\end{table*}

\begin{table*}[htp]
\begin{center}
\caption{References for Table \ref{tab:clusters}.}
\label{tab:opticalref}
\begin{tabular}{l l l l}
\hline\hline
\textbf{Symbol}     &    \textbf{Reference}       & \textbf{Symbol}     &      \textbf{Reference}  \\
\hline
62       &    \citet{Geller1984}        &77     &   \citet{Wu1995}          \\
63       &    \citet{Chernomordik1995}  &78     &   \citet{Condon1990}      \\
64       &    \citet{West1991}          &79     &   \citet{Kim1991}         \\
65       &    \citet{Dobrzycki2002}     &80     &   \citet{ODea1996}        \\
66       &    \citet{Edge1995}          &81     &   \citet{Hintzen1984}     \\
67       &    \citet{Ryle1968}          &82     &   \citet{Nair1997}        \\
68       &    \citet{Cristiani1997}     &83     &   \citet{White1999}       \\
69       &    \citet{Zwicky1966}        &84     &   \citet{Golev1998}       \\
70       &    \citet{Lipovka2002}       &85     &   \citet{Smail2003}       \\
71      &    \citet{Ledlow1995}         &86     &   \citet{Hintzen1983}     \\
72      &    \citet{Zwicky1968}         &87     &   \citet{Andernach1988}   \\
73      &    \citet{Tyson1986}          &88     &   \citet{Hutchings1998}   \\
74      &    \citet{Dunn2005}           &89     &   \citet{Miller2002}      \\
75      &    \citet{Hill1991}           &90     &   \citet{Gehren1984}      \\
76      &    \citet{Maoz1995}           &       &                           \\
\hline
\end{tabular}
\end{center}
\end{table*}

\begin{table*}[htp]
\begin{center}
\caption{CJF sources observed to have a companion and sources showing evidence of an ongoing interaction with another system. Columns (1)-(4) as previously, Col. (5) gives the optical magnitude, Col. (6) gives cluster information for comparison (Table \ref{tab:clusters}), Cols. (7) and (8) give references in regard to companion galaxies and their separation respectively, and Col. (9) gives the references in regard to interacting systems. By a star, we indicate ambiguous evidence.}
\label{tab:tidal}
\begin{tabular}{l l l l l l l l l}
\hline\hline
\textbf{Source}	&	\textbf{Other Name}		 &\textbf{Type}	 &\textbf{z}	       &\textbf{m}		&\textbf{Cluster} &\textbf{Companion} &\textbf{Separation} &\textbf{Interaction}\\
	    &				     &	     &		   &(mag)	& & &(kpc) &\\
\hline
0018+729&				     &G	 &0.821	   &		&	   	&	    	&		&91		          \\
0108+388&	OC 314			 &G	 &0.669	   &22		&	   	&	    	&		&91		          \\
0316+413&	3C 84			 &G	 &0.018	   &11.9	&67     &92         &$\sim600$&92             \\
0402+379&	4C 37.11		 &G	 &0.055	   &18.5	&	   	&2	    	&29		&2               \\
0710+439&				     &G	 &0.518	   &20.4	&62	   	&93*	    &53		&93               \\
0804+499&	OJ508			 &Q  &1.432	   &17.5	&	   	&        	&		&97               \\
0831+557&	4C 55.16		 &G	 &0.240	   &17.5	&74	   	&93*	    &81		&93*              \\
0954+658&				     &BL &0.368	   &16.7	&	   	&94     	&474	&		          \\
1031+567&	COINS J1035+5628 &G	 &0.460	   &20.2	&91	   	&80	    	&65		&49                \\
1146+596&	NGC3894			 &G	 &0.011    &13		&84	   	&	    	&		&98               \\
1254+571&	Mrk 231			 &G	 &0.042    &13.6	&85     &37         &361	&37               \\
1418+546&	OQ530			 &BL &0.151	   &15.5	&82	   	&	    	&		&99               \\
1641+399&	3C345			 &Q  &0.595	   &16		&88	   	&22,95   	&47		&   	          \\
1652+398&	Mkn 501   		 &BL &0.034    &13.8	&89	   	&	    	&		&99               \\
1807+698&	3C 371.0		 &BL &0.051	   &14.8	&82	   	&94         &81		&99               \\
1823+568&   4C 56.27         &BL &0.664    &18.4    &       &61         &46     &61               \\
1843+356&	COINS J1845+3541 &G	 &0.764	   &21.9	&	   	&80*	    &129	&91               \\
1946+708&				     &G	 &0.101	   &16.1	&	   	&96	    	&67		&96               \\
2021+614&	OW 637			 &Q  &0.227	   &19		&80*   	&80*        &36		&	              \\
2352+495&	OZ 488			 &G	 &0.237	   &19		&80	   	&80*	    &47		&                 \\
\hline
\multicolumn{2}{l}{91. \citet{Stanghellini1993}}&\multicolumn{3}{l}{92. \citet{Colina1990}} &\multicolumn{2}{l}{93. \citet{Hutchings1988}} \\
\multicolumn{2}{l}{94. \citet{Stickel1993}} &\multicolumn{3}{l}{95. \citet{Kirhakos1999}} &\multicolumn{2}{l}{96. \citet{Perlman2001}} \\
\multicolumn{2}{l}{97. \citet{Hutchings1999}} &\multicolumn{3}{l}{98. \citet{Pustilnik2001}} &\multicolumn{2}{l}{99. \citet{Pursimo2002}} \\
\end{tabular}
\end{center}
\end{table*}

We also investigate the CJF sample for the number of sources that belong to a cluster or a group of galaxies. Because of the large redshift span of the CJF, it is not possible to have cluster information for sources at high z ($z > 1.5$, e.g., \citealt{Brodwin2008}; \citealt{Blakeslee2003}).  We find 54 sources (18.4\% of the sample) belonging to a cluster or group of galaxies (see Table \ref{tab:clusters}). For $z < 1$, we calculate that 25.0\% of the QSOs ($z_{avg}=0.60$), 33.3\% of the radio galaxies ($z_{avg}=0.27$), and 29.4\% of the BL Lac objects ($z_{avg}=0.25$) are found in clusters. We note that radio galaxies appear to be more often members of clusters. \citet{Hill1991} find a similar trend in their redshift range (e.g., see Fig. 11 of \citealt{Hill1991}).
Twenty-two sources (7.5\% of the sample) show variable fluxes in the optical (intra-day variability has not been taken into account). Some of these sources have extensive enough lightcurves to calculate the timescales of this variability (see Table \ref{tab:multivar}).

\section{Analysis and results}
\label{sec:results}

\subsection{Mergers in the CJF}

Assuming that activity (whether nuclear or star-formation) in galaxies is indeed triggered by mergers, we expect to observe morphological distortions in the optical. If these structures are observed, the system must have recently (100-200 Myr, \citealt{Lotz2008}) undergone a merger. A typical merger lifetime is of the order of a few billion years. Using simulations, \citet{Lotz2008}, find that the peak in the morphological disturbances of the merging system happens just after the first pass of the two galaxies and after the actual merging of the nuclei of the merging galaxies. In-between, the system does not exhibit any quantitatively appreciable asymmetry. Merger-induced star formation can be used as evidence of merging systems in this intermediate phase.

We find 16 sources (5.5\% of the sample, 11 radio galaxies, 4 BL Lac objects, and 1 quasar) exhibiting distortions in the optical and infrared (see Table \ref{tab:tidal}). We find 14 sources (4.8\% of the sample, 9 RG, 3 BL, and 2 Q) with one or more close companions (Table \ref{tab:tidal}). The percentage of distorted sources or sources with companions is in good agreement with \citet{Hutchings1983} and \citet{Surace1998}, who find that, for $z<0.4$ a 30\% of quasars show such properties (28.6\% for the CJF). This underlines the ubiquity of these morphologically distorted systems and the importance of mergers in the AGN phenomenon. The angle to the line of sight assumed for radio galaxies (\citealt{Urry1995}) allows the investigation of the host galaxy, unlike in the case of BL Lacs and quasars, where the jet dominates the emission. This may explain the higher percentage of radio galaxies in the distorted sources. Moreover, their lower average redshift compared to the quasar sample, contributes to the above effect. Finally, with lower nuclear luminosity, radio galaxy cores do not outshine their host galaxies, thus facilitating these morphological studies.

Given the small number of CJF sources observed in the far-IR, information about the star-formation rates of these objects are scarce. Nevertheless, we find four sources that have signs of starburst activity (0316+413, \citealt{Richer1993}; 1254+571, \citealt{Richards2005}; 1652+398, 1807+698, \citealt{Condon2002}) and one additional candidate (0248+430, \citealt{Dudik2005}).

Finally, since we find one ULIRG (Mrk 231) in our sample, we can infer a percentage for the content of ULIRGs in the CJF sample. Counting only the IR detected sources, we have a 1.0\% of the sample exhibiting very high luminosities in the far-IR ($>10^{12}L_{sol}$. This number is a lower limit, given that not all CJF sources have been observed in the far-IR or are observable at their respective redshifts.

\subsection{Luminosity correlations}

\begin{figure*}
\begin{center}
  \includegraphics[width=0.90\textwidth,angle=0]{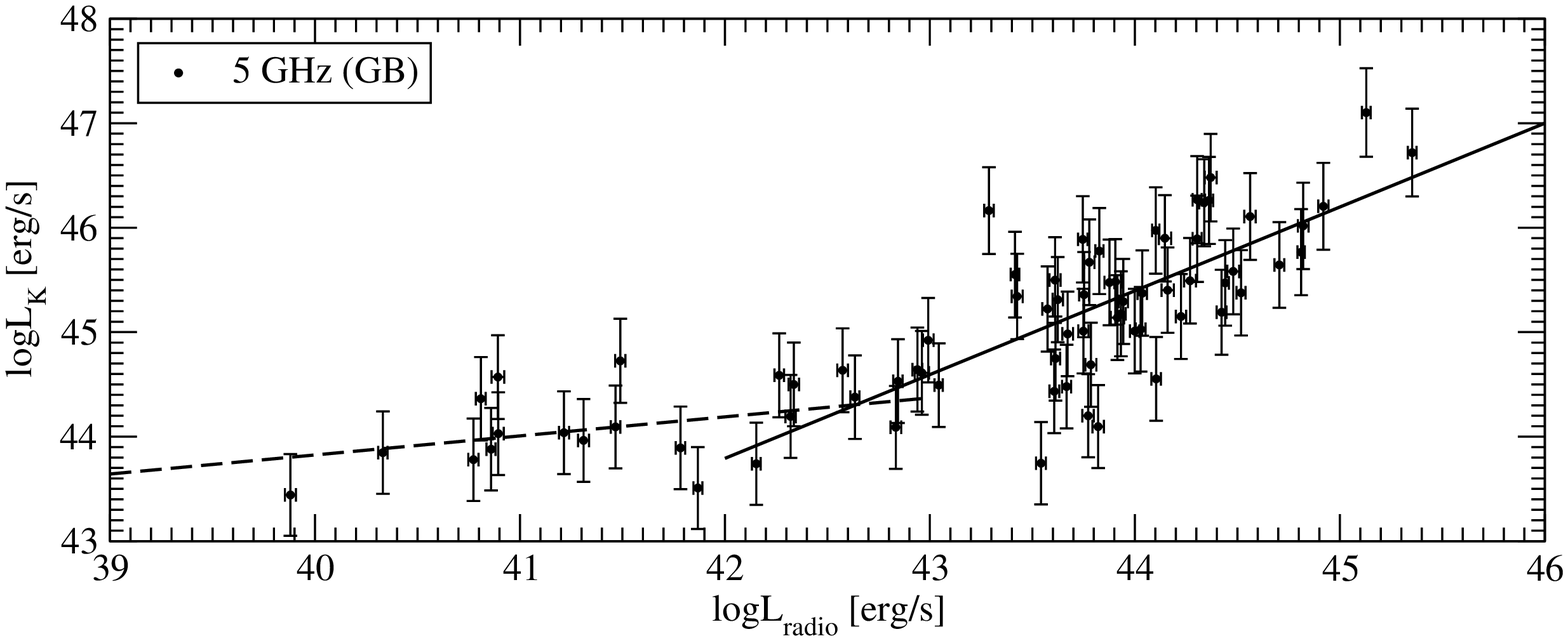}
  \caption{Luminosities in the near-infrared as a function of the radio luminosity at 5 GHz, from single dish observations with the Green Bank radio telescope (data from \citealt{Gregory1991}). Linear regression fits for the low radio luminosity (dashed line) and high radio luminosity (continuous line) CJF sources are shown.}
  \label{fig:radio_core_luminosities}
\end{center}
\end{figure*}
In Fig. \ref{fig:radio_core_luminosities}, we plot the near-infrared (K band) luminosity against the K-corrected radio luminosity obtained by single dish observations at 5 GHz.

By measuring the infrared luminosity of our sources, we probe the stellar light of the active galaxies, as well as radiation from the dust and gas component of the sources. We are interested in the relation of this thermal component to the non-thermal one, given by the radio luminosity of the AGN.  We find a broken correlation between these two quantities. The lower luminosity sources appear to have a near constant NIR luminosity independent of the radio core luminosity. A standard linear regression infers a slope of $0.18\pm0.12$ with a correlation coefficient of $0.35$. For higher luminosities, a linear dependence is found. A slope of $0.80\pm0.11$ with a correlation coefficient of $0.79$ is given by standard linear regression. Following \citet{Browne1987}, we examine in addition the correlation between $L_{radio}/L_{NIR}$ and both $L_{radio}$ and z. We find a stronger correlation with $L_{radio}$ (correlation coefficient of 0.788) than to z (correlation coefficient of 0.498). This leads us to believe that the correlation is not an artifact of luminosity evolution. The large scatter of the data is expected, because of significant variability of active galaxies and the radio and infrared observations not being contemporaneous.

\citet{Rieke1978} was first to confirm a correlation between NIR and radio continuum for a sample of Seyfert galaxies. This indicates that the infrared output of Seyfert galaxies is dominated by reradiation from dust in the host galaxy. In this context, we find that higher luminosity sources are more dominated by their AGN component (as traced by the radio luminosity), than lower luminosity sources, for which an excess NIR luminosity is found. This excess is measured relative to the luminosity predicted by the linear dependency seen for higher luminosity sources and indicates that there is significant infrared output from star formation regions that dominates over nuclear infrared emission. Therefore, higher redshift detected (effectively, higher luminosity) sources are less likely to be found in a transitionary evolutionary stage, since their spectrum is exclusively dominated by their nucleus, effectively outshining and hiding any effects of an ongoing merger.

\subsection{AGN near-IR colors}

\begin{figure*}[htp]
\begin{center}
  \includegraphics[width=0.50\textwidth,angle=-90]{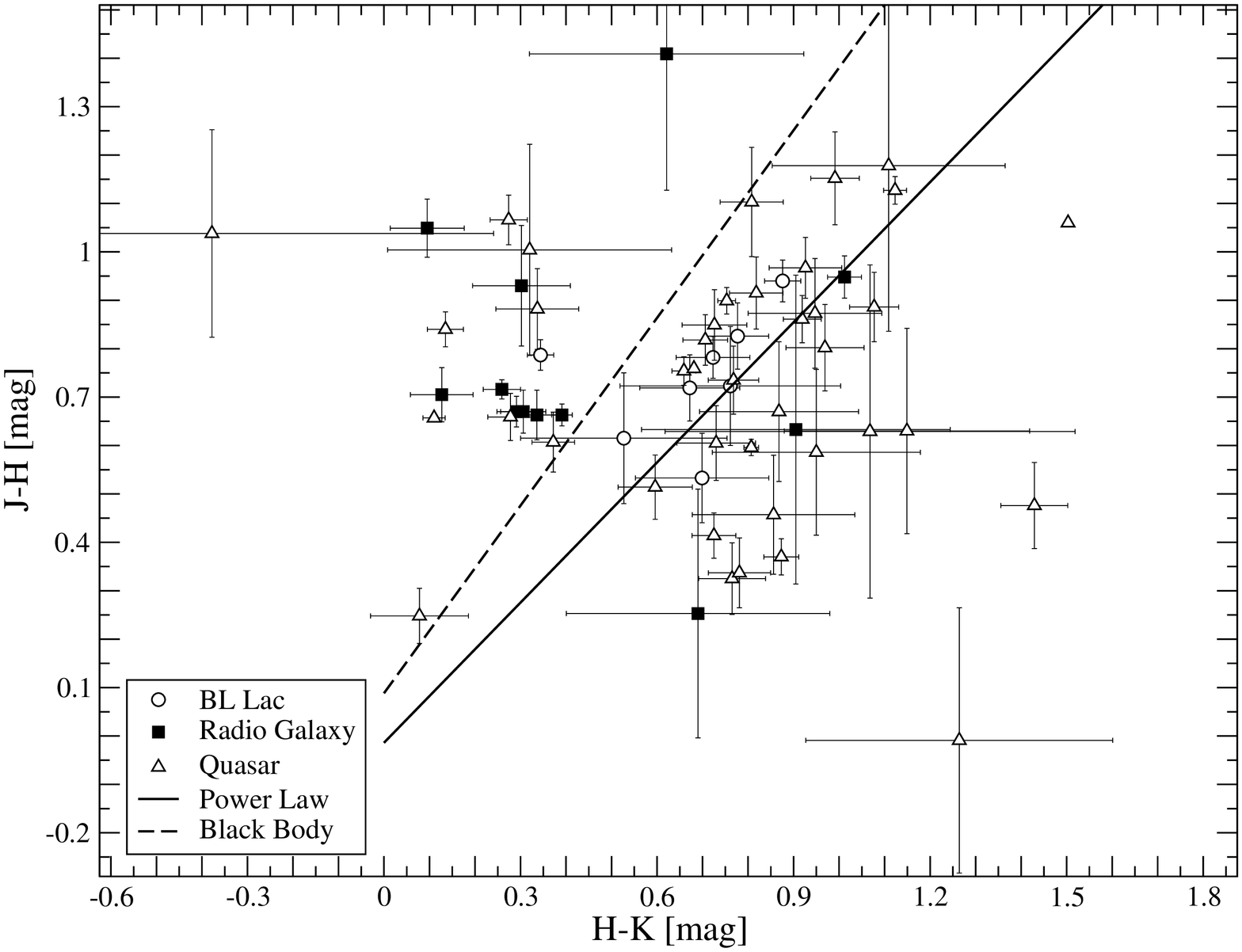}
  \caption{Near-infrared color-color diagram for all CJF sources with available information. We plot the lines for a power-law emitting body (continuous line) and for a black body (dashed line). We note blazars with open symbols (circles for BL Lac objects and triangles for the quasars) and radio galaxies with filled ones (see text for details).}
  \label{fig:color_diagram_types_all}
\end{center}
\end{figure*}

In the same context as above, we are interested in investigating the interplay between the star-formation and nuclear component in the sources of our sample. For this reason, we calculate the near-infrared colors of the sources with available data, after correcting all magnitudes (HJK bands) for Galactic extinction (from \citealt{Schlegel1998}). From these, we construct the near-IR color-color diagram seen in Fig. \ref{fig:color_diagram_types_all}. The lines for a black body and a power-law emitting body divide the color-color parameter space into three regions. Sources situated above the black-body line are predominantly thermal emitters. Sources beneath the power-law line are dominated by non-thermal emitting mechanisms. For sources between the two lines, both components are important.

In Fig. \ref{fig:color_diagram_types_all}, we separated blazars from radio galaxies according to the different nature of these objects. We anticipate that radio galaxies are dominated by emission originating in their host galaxies. Indeed 8 out of the 11 radio galaxies depicted in Fig. \ref{fig:color_diagram_types_all} are situated above the black-body line. Radio galaxies appear significantly redder in both J-H and H-K colors with $(J-H)_{avg,RG}=2.2\pm1.1$ and $(H-K)_{avg,RG}=1.8\pm1.1$. With the exception of nine sources, the remaining blazars are situated beneath that same line, most being situated around and beneath the power-law line.

We are particularly interested in the sources found in the space between the two lines, hence showing transitory properties from a thermal emission dominated state to a non-thermal dominated one. Indeed blazars in this transition area, with an average H-K color of $(H-K)_{avg}=0.814\pm0.027$, are considerably redder than their equivalents found above the black-body line ($(H-K)_{avg}=0.19\pm0.05$), sign of an excess of dust leading to the observed reddening.

In the previous section, we inferred a possible connection between multiwavelength variability, binary black holes, and hence post-merger or merger-remnant systems. We test this connection using the near-IR color properties of our sources. We divide our sources into three groups, as defined by the two lines seen in Fig. \ref{fig:color_diagram_types_all}. We find that 42.9\% of the sources situated between the black-body and power-law lines, effectively sources assumed in a transition from a merger (a system that is dominated by its thermal emission) to an active galaxy (a system that is dominated by its non-thermal emission), are variable at different wavelengths. Only 11.1\% and 10.5\% of the sources above the black-body line and below the power-law line, respectively, show variability.

\section{Tracing the evolution in the CJF}
\label{sec:tracing}

One aim of this paper is to study the possible evolutionary stages of the AGN in the CJF sample. Different subsamples of the CJF, and the individual sources within them, can probe these different evolutionary stages. Mergers, starburst activity, and BBH systems are integral parts of this study.

One of the most prominent examples supporting the link between starburst and BBH systems is NGC 6240, one of few sources directly observed to have a binary core (\citealt{Komossa2003}). This system is an ULIRG, hosts two AGN, and also clearly exhibits ongoing starburst activity, making it an archetype of the evolutionary scenario discussed here.

By selecting sources with a companion, one can investigate the earliest stages of the merging process. Accordingly, by selecting the sources with disturbed morphologies and large infrared fluxes, we probe an intermediate phase (e.g., Mrk 231). Finally, sources with relaxed morphologies but almost periodic variability at multiple wavelengths are likely to represent systems in which an assumed BBH has sunk to the center of the system.

\citet{Lotz2008} conducted simulations of equal-mass gas-rich mergers, classifying the sources into six distinct merging phases: pre-merger, first pass, maximal separation, merger, post-merger, and remnant. Using this classification scheme for our sources, we classify objects into pre and post-merger stages.

The incompleteness of the information available about our sample does not allow us to make unambiguous claims for the classification of some sources (especially when differentiating between for example either pre-merger and maximal separation, or post-merger and remnant classes). Among the distorted sources, those with companions  can be classified as either a pre-merger or maximal separation phase, while those with no companions can be categorized as being in either the first pass, the post-merger, or remnant phase. For the last group, this translates, roughly, to an age of 2-4 Gyr after the initial approach of the progenitor systems. Exceptions to the above constitute four sources (3C 84, Mrk 231, 3C 371, and Mrk 501) that exhibit both distorted morphologies and considerable starburst activity (e.g., \citealt{Richards2005}; \citealt{Condon2002}). These characteristics put them in the merger phase and would constrain their age to be $<2$ Gyr after the initial approach.

\begin{table*}
\begin{center}
\caption{CJF sources categorized in different evolutionary phases of AGN (following \citealt{Lotz2008}, for details see text). Columns (1)-(4) as in previous tables, Col. (5) provides information about companions, Col. (6) gives morphology information, Col. (7) indicates the number of wavelength regimes the source has been observed and detected as variable, and Col. (8) describes the starburst activity.  For references, see previous tables.}
\label{tab:evolution_candidates}
\begin{tabular}{l l l l l l l l}
\hline\hline
\textbf{Source}	      &	\textbf{Other Name}	&\textbf{Type}	&\textbf{z}	&\textbf{Companion} &\textbf{Morphology}	 &\textbf{Variability}	  &\textbf{Starburst}	 \\
\hline
\multicolumn{8}{l}{\textbf{Pre - merger}}	 \\
\hline
0954+658	&		    &BL Lac	&0.368	&yes	  	&	&	2	  &             \\
2021+614	&OW 637		&Q	    &0.227	&yes	  	&	&		  &             \\
2352+495	&OZ 488		&G	    &0.237	&yes	  	&	&		  &             \\
\hline
\multicolumn{8}{l}{\textbf{First Pass}}	     \\
\hline
0018+729	&		    &G	    &0.821	&	   &disturbed	&		  &             \\
0108+388	&OC 314		&G	    &0.669	&	   &disturbed	&		  &          	\\
0804+499	&OJ508		&Q	    &1.432	&	   &tidal tail	&	1	  &             \\
1146+596	&NGC3894	&G/LINER&0.011	&      &interacting	&		  &             \\
\hline
\multicolumn{8}{l}{\textbf{Max Separation}}	          \\
\hline
0402+379	&4C 37.11	&G/Sy	&0.055	&yes	   &interacting	          &	1	  &             \\
0710+439	&OI+417		&G/Sy	&0.518	&yes	   &disturbed	          &		  &             \\
0831+557	&4C +55.16	&G/LINER&0.240	&yes	   &disturbed	          &		  &             \\
1031+567	&		    &G/Sy2  &0.460	&yes       &disturbed	          &		  &             \\
1823+568    &4C 56.27   &BL Lac &0.664  &yes       &non-relaxed isophotes &       &             \\
1843+356	&		    &G	    &0.764	&yes	   &disturbed	          &		  &             \\
1946+708	&		    &G	    &0.101	&yes	   &non-relaxed isophotes &	      &             \\
\hline
\multicolumn{8}{l}{\textbf{Merger}}	          \\
\hline
0248+430	&		    &Q	    &1.310	&      &	                    &		  &possible \\
0316+413	&3C 84		&G/Sy2	&0.018	&yes   &twisted isophotes	    &	3	  &yes   	\\
1254+571	&Mrk 231	&Q	    &0.042  &yes   &disturbed	            &	3	  &yes   	\\
1652+398	&Mrk 501	&BL Lac	&0.034  &      &extra nuclear component	&	5	  &yes   	\\
1807+698	&3C 371.0	&BL Lac	&0.051	&yes   &interacting	            &	2	  &yes   	\\
\hline
\multicolumn{8}{l}{\textbf{BBH / Post-merger}}	            \\   	
\hline
0219+428	&3C066A		&BL Lac	&0.444	&	  	&	                    &	3	  &       	    \\
0716+714	&		    &BL Lac	&	    &	  	&	                    &	5	  &             \\
0814+425	&OJ425		&BL Lac	&0.245	&	  	&	                    &	3	  &             \\
0923+392	&4C 39.25	&Q	    &0.699	&	  	&	                    &	3	  &             \\
1101+384	&Mrk 421	&BL Lac	&0.031	&	  	&	                    &	5	  &             \\
1418+546	&OQ530		&BL Lac	&0.151	&	    &extra nuclear component&	3	  &      	    \\
1633+382	&4C 38.41	&Q	    &1.807	&	  	&	                    &	3	  &             \\
1641+399	&3C345		&Q	    &0.595	&yes	&	                    &	4	  &             \\
2200+420	&BL Lac		&BL Lac	&0.069	&	  	&	                    &	4     &             \\
\hline
\end{tabular}
\end{center}
\end{table*}
In Table \ref{tab:evolution_candidates} we indicate the 27 most suitable representatives of different evolutionary phases (following \citealt{Lotz2008}). Sources with companions but no sign of disturbed morphologies, are selected as pre-merger systems. Sources with disturbed morphologies but no companions and no apparent starburst activity, are assigned to the first pass phase when the two systems are at a minimum separation for the first time. The next phase corresponds to that of the maximal separation between the merging systems. For this phase, sources with disturbed morphologies and detected companions are selected. Sources identified as undergoing the first pass may also be merger remnants. More information is required to make this distinction.

In the context of BBH systems, the last two categories shown in Table \ref{tab:evolution_candidates} are the most interesting, as we expect the SMBH at the centers of the merging systems to be close enough for the gravitational pull to bind them together. For the merger phase, we select sources that have disturbed morphologies and also show signs of ongoing starburst activity (similar to Mrk 231). For the final category we select sources that appear extremely variable across their spectrum (variable in at least 3 different wavelength regimes) in an almost periodic manner. These sources do not show signs of ongoing interaction, merging (apart from 1418+546, see \emph{Individual Sources}), or starburst activity. These are probably BBH systems. In the context of the above categorization, we can identify these sources as possible post-merger systems.

\begin{figure}
\begin{center}
  \includegraphics[width=0.35\textwidth,angle=-90]{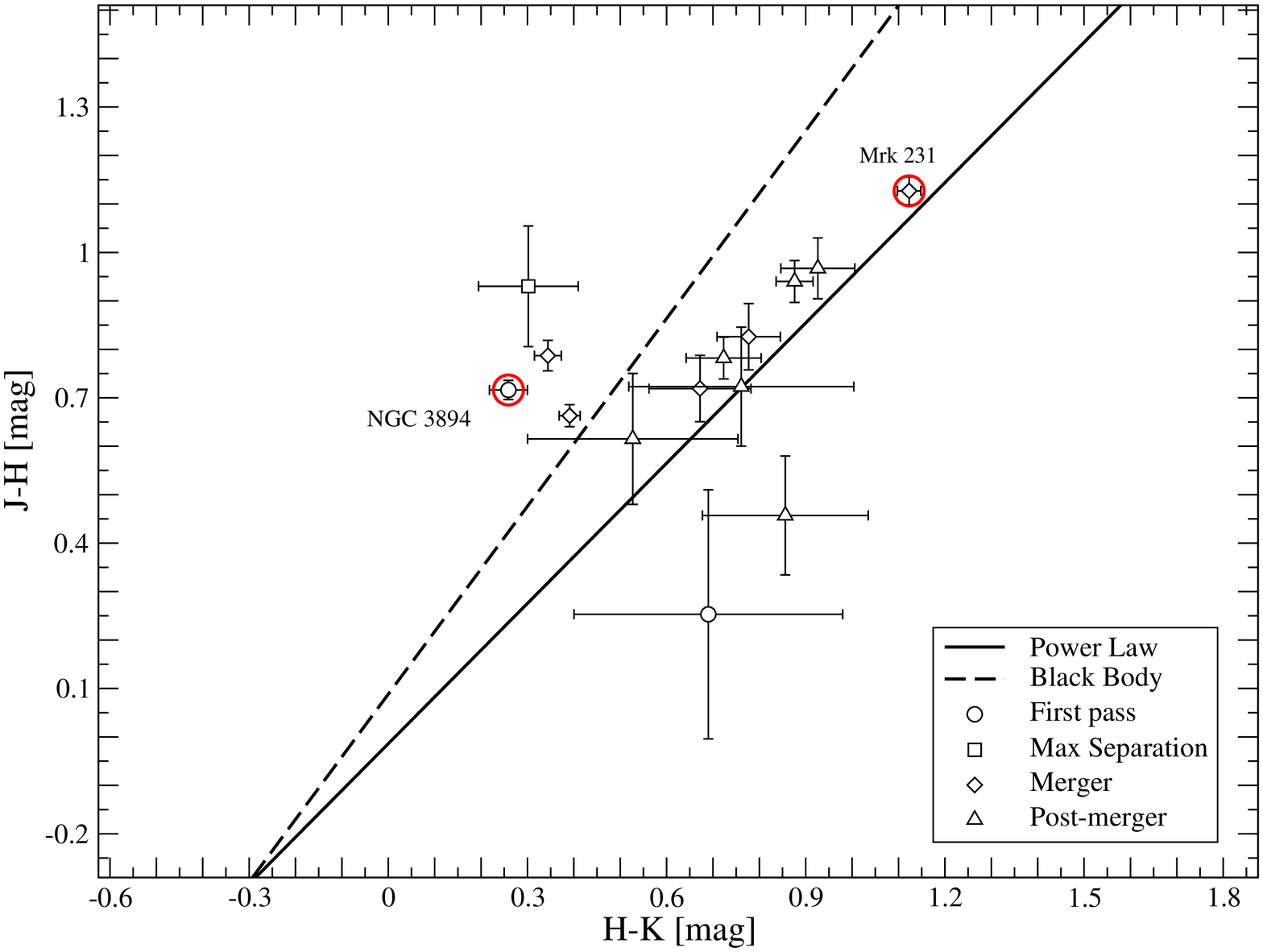}
  \caption{Near-infrared color-color diagram for the sources included in Table \ref{tab:evolution_candidates} with available information. We plot the lines for a power-law emitting body (continuous line) and for a black body (dashed line). We note two objects (Mrk 231, NGC 3894) of interest due to their transitionary nature (see text for details).}
  \label{fig:color_diagram_candidates}
\end{center}
\end{figure}

At least three of the sources in Table \ref{tab:evolution_candidates} (0954+658, 0108+388, and 0248+430) cannot be unambiguously categorized as one evolutionary phase, as they exhibit properties common to more than one stages of evolution. Sources selected in the last two categories (merger and post-merger) are the most likely BBH candidates in the CJF sample. We plot their NIR colors in Fig. \ref{fig:color_diagram_candidates}. As expected, most candidate sources are found to be above the power-law line, while most post-merger classified sources (selected solely by their variability properties) are found just above the power-law line, in the transitory area of the color-color diagram. This provides us with an additional selection tool for transitory systems (see Sect. \ref{sec:individual}).

\section{Discussion}
\label{sec:discussion}

In previous sections, we have presented the accumulated literature information available for the CJF sources and a sample of CJF sources that trace AGN evolution, indicating the most likely BBH candidates among them. We now investigate the criteria used to identify candidates as BBH systems. It is interesting to discuss which subset of these criteria can provide us with a robust enough argument for the existence of a BBH.

\subsection{Anticipated AGN properties}
Before we proceed with the discussion, we are interested in assessing the completeness of the information for the CJF sample presented here, as these were collected from the available literature and are not the result of a uniform observational campaign. To achieve this goal, we create two subsamples of CJF sources. For the first one, we selected sources with available near-infrared information, as this information was primarily presented and analyzed in this paper. In addition, we excluded sources without redshift information. We identified 74 of these sources, the number of sources being explicitly controlled by the availability of infrared information (see Table \ref{tab:infrared}). For the second subsample, we selected those CJF sources that have been observed and have available information across the whole spectrum (radio to $\gamma$-rays). We found only 10 of these sources, the number of sources mainly being regulated by availability of $\gamma$-ray information (see Table \ref{tab:cjfsubsample}).
\begin{center}
\begin{table}[h]
\caption{The CJF source with available information across the electromagnetic spectrum (radio, near-infrared, optical, X-ray, and $\gamma$-ray).}
\label{tab:cjfsubsample}
\begin{tabular}{l l l l}
\hline\hline
\textbf{Source}		&\textbf{Other Name}	&\textbf{Type}	&\textbf{Redshift}	\\
\hline
0133+476	&OC 457		&HPQ	&0.859          \\
0219+428	&3C066A		&BL Lac	&0.444          \\
0316+413	&3C 84		&G	&0.018          \\
0650+453	&		&Q	&0.933          \\
0814+425	&OJ425		&BL Lac	&0.53           \\
0836+710	&4C 71.07	&LPQ	&2.18           \\
1101+384	&Mrk 421	&BL Lac	&0.0308         \\
1633+382	&4C 38.41	&LPQ	&1.807          \\
1652+398	&Mrk 501	&BL Lac	&0.03366        \\
1803+784	&		&BL Lac	&0.6797         \\
2200+420	&BL Lac		&BL Lac	&0.0686         \\
\hline
\end{tabular}
\end{table}
\end{center}
\subsubsection{Near-IR subsample}
This subsample, having been selected on the basis of detection in the near-IR, is expected to be more sensitive to infrared bright sources and consequently have a higher probability of containing sources associated with merger events. This bias should be accounted for. We begin by investigating whether this subsample is differs fundamentally from the CJF sample from which it was selected. The subsample has an average redshift of $z_{avg,sub}=0.864$, effectively probing more nearby sources than the CJF ($z_{avg, CJF}=1.254$). We also check the radio spectral index $\alpha^{4800}_{1400}$, a measure of the compactness of the sources. The subsample has lower average and median values of the index than the whole sample. This indicates that this subsample contains less compact objects. Finally we check the range of luminosities for which almost complete information for the CJF is available (in radio, optical, and X-ray) for both the subsample and the CJF as a whole. In both samples, we find that the ranges are the same for all three different wavelengths. We conclude that this subsample probably contains sources that are closer and more extended but is otherwise quite similar to the CJF.

In the context of merger-driven evolution, we find that 18.9\% of the sources in this subsample, exhibit morphological distortions and starburst activity. Eleven sources (14.9\%) are found to have companions.  Compared to the CJF sample in total (5.5\% of sources exhibit distortions; 4.8\% have companions), merger effects appear to be more prominent in this subsample, as expected. However, the more than three times higher percentage of distorted sources and sources with companions than the CJF as a whole indicates that the true number of these sources might be higher than currently believed.

We are also interested in the variability properties of this subsample: 22.7\% of the sources are variable in the radio, 13.3\% in the infrared, 18.7\% in the optical, 6.7\% in the X-rays, and 12.0\% in the $\gamma$-rays. For most of these sources and wavelength regimes (with the exception of infrared and $\gamma$-rays), periodicities have been found in their lightcurves (see Tables \ref{tab:multivar} and \ref{tab:multivarref} for timescales and references). Compared to the total CJF sample, these sources appear to be significantly more variable. This is an unexpected result, as nothing in our selection criteria and the properties of this sub-sample implies a bias towards more variable sources.

\subsubsection{Multiwavelength subsample}
This second subsample was selected to assess how our results (concerning the importance and relevance of merging events and variability for active galaxies) is affected by the availability of multiwavelength information (or lack thereof). Given the very small number of the sources included in this subsample, a statistical comparison to the whole sample is difficult. The average redshift of this subsample is $z_{avg,sub}=0.689$, and the subsample consists almost entirely of blazars (with the exception of 3C 84).

In the context of merger-driven evolution, we find that for this subsample, 17.6\% of the sources (all of which are at z$<$0.05) exhibit morphological distortions and starburst activity. One source (9.1\%) is found to have companions.  Compared to the CJF sample in total, merger effects appear to be more prominent in this sub-sample. One must again take into account that the average redshift of the CJF sample is $z_{avg, CJF}=1.254$, much higher than this subsample. However, the more than three times higher percentage of distorted sources compared to the CJF as a whole indicates that the true number of these sources is higher than currently believed.

The variability properties of this subsample are: 81.8\% of the sources are variable in the radio, 45.4\% in the infrared, 72.7\% in the optical, 27.3\% in the X-rays, and 81.8\% in the $\gamma$-rays (see Tables \ref{tab:multivar} and \ref{tab:multivarref} for timescales and references, respectively). Compared to the total CJF sample, these sources appear to be highly variable, in an almost periodical manner. This can be explained as a bias effect, as sources that are highly variable in one waveband are more likely to be targeted for observations in other wavelengths. Alternatively, one can argue that (periodical) variability is an ubiquitous phenomenon for a large portion of active galaxies. The latter is also supported by the high percentages of variable sources detected for the first sub-sample. Multi-wavelength variability, however, can only be detected by extensive (and therefore time-costly) monitoring of a source, thus effectively reducing the number of known sources with such properties.

For both subsamples, we find rather high percentages of sources with signs of past, present, or future merger events, as well as sources with variable fluxes at different wavelengths. Although the criteria used to select these two subsamples were different, the resulting percentage of sources with merger signs are the same for both subsamples, significantly higher than the percentage for the whole CJF. This implies that because of the incompleteness of the information available to us (due to both a lack of observations and instrumental limitations), the percentage of active galaxies found to be linked to a merger event in our sample is only a portion of the true number, thus is only a lower limit. If the percentage calculated for the two subsamples were closer to the true percentage, we would expect approximately 50 CJF sources to have been found taking part in a merger event and 31 sources having close companions. From the available information, we have found 15 and 11 sources, respectively.

\subsection{Periodicities in the lightcurves}

Having concluded that this statistical analysis of the CJF can provide us with only lower limits for, e.g., the percentage of mergers in our sample of active galaxies, we can now discuss the criteria used to identify BBH systems.

Periodicities in the lightcurves can be linked to a periodic process at the center of an AGN, presumably a binary system with a quasi-stable orbit that produces periodical perturbations in the accretion disk and the jet of the AGN (e.g., \citealt{Lobanov2005}; \citealt{Qian2007}; \citealt{Rieger2004}; \citealt{Roland2008}; Britzen et al. 2009, in press). However, this is not the only possibility.

Precession of the jet has been employed to explain a variety of phenomena including the variability that AGN show at different wavelengths (e.g., \citealt{Caproni2004}; \citealt{Roland1994}). An intrinsically helical jet can also exhibit the same morphological and kinematical results (for a study of the different kind of helicities in AGN jets, see \citealt{Steffen1997}). Helicity in AGN jets may be the result of internal Kelvin-Helmholtz instabilities and in particular the $m=2$ helical mode of the K-H instabilities (for a 2-D treatment of the problem see e.g., \citealt{Baty2003}; alternatively, see \citealt{Perucho2006}; \citealt{Meier2006}).

Doppler beaming may also cause the observed variability in the lightcurves of AGN. Camenzind \& Krockenberger (1992) first described the so-called lighthouse effect. This model has been successfully applied to different sources (e.g., 3C 273, \citealt{Camenzind1992}; 3C 345, \citealt{Schramm1993}).

Different physical phenomena are described by different characteristic times. The BBH and precessing jets can describe periodicities of the order of $~10$ years (see references above), whereas lighthouse models tend to explain the intra-day variability, also observed in many blazars (for a review, see \citealt{Wagner1995}). Intra-day variability can also be explained by interstellar scintillation (for a review, see e.g., \citealt{Rickett1990}; also see \citealt{Melrose1994}), in addition to BBH models (\citealt{Roland2009}).

Correlated variability at different wavelengths implies that the same population of particles produces this variability, thus pointing to an intrinsic effect, rather than, for example, a turbulent screen between the observer and the observed source. One of the most prominent sources displaying correlated variability at different wavelengths is 0716+714 (radio and optical) (\citealt{Quirrenbach1991}; \citealt{Wagner1996}). The source OJ287 is another famous example, that has correlated variability in the radio, optical, and infrared (see e.g., \citealt{Fan1999}). OJ287 is a prominent candidate BBH system (\citealt{Sillanpaa1988}; \citealt{Lehto1996}; \citealt{Pietil1998}; \citealt{Valtonen1999}; \citealt{Liu2002}; \citealt{Valtonen2007}).

We categorized possible post-merger candidates as sources that exhibit almost periodic variabilities in their lightcurves (Table \ref{tab:evolution_candidates}), assuming that a BBH causes the variability at different wavelengths. Indeed we find that variability is connected to transitory systems as they are defined on the near-IR color-color diagram. In Table \ref{tab:multivar}, we indicate the variability timescales where available. Out of 9 post-merger classified sources, 4 sources (0716+714, 1101+384, 1641+399, 220+420) show almost periodic variability with similar timescales ($\sim10$ years) in the radio and optical. These are candidate hard BBH systems. More detailed investigation of their radio jet morphology may provide us with additional information.

\subsection{Infrared emission, merging scenarios, and starbursts}

The infrared emission from an AGN consists of different components, coming from different parts of the nucleus. Distinguishing and separating these constituents (accretion disk, thermal, and jet, non thermal) is not straightforward. Both components would be affected by a BBH system, since both would precess under the gravitational influence of the binary, although possibly on different timescales.

In light of the merger-driven evolutionary scenario, we investigate whether there is a connection between an excess in the infrared regime and sources with companions or sources that exhibit disrupted morphologies. For 16 sources with peculiar morphologies, we calculate an average infrared (K band) luminosity $log(\nu L_{\nu})_{NIR,avg}=44.17\pm0.14\:erg/s$ for an average redshift of $z_{avg}=0.41$. We also calculate the average luminosity in the NIR for sources that are found to have companions $log(\nu L_{\nu})_{NIR,avg}=44.61\pm0.21\:erg/s$, for 11 sources with a $z_{avg}=0.34$. Defining a subsample of the CJF sources with similar average and median redshifts ($z_{avg}=0.38$), we find $log(\nu L_{\nu})_{NIR,avg}=44.5\pm0.1\:erg/s$. Sources with disturbed morphologies are of below average near-infrared luminosity, in contrast to sources with companions that are on average and within the errors, equally luminous to the rest of the sample. The available data do not show a prominent near-IR excess for AGN with perturbed morphologies as would be expected. The lack of mid- and far-infrared data does not allow us to extend this over the whole infrared range.

Mrk 231 is an ULIRG, classified as a broad line quasar (BLQ) and a type 1 Seyfert. The system shows signs of an ongoing interaction and merger (\citealt{Canalizo2001}; \citealt{Petrosian2002}). The galaxy hosts an AGN, with a Seyfert 1 spectrum, but is however also classified as a quasar. The source exhibits strong starburst activity (\citealt{Richards2005}), which indicates that it has undergone a merger. The source has been studied in X-rays, where it exhibits extended emission, in addition to a hard spectrum central source that remains unresolved and is off-center to the extended X-ray halo. An off-nuclear source has been detected at X-rays at a distance of $6''$ (\citealt{Gallagher2002}). Mrk 231 has been studied for variability at different wavelengths. It has been found to be variable in the X-rays (\citealt{Gallagher2002}), optical (in the absorption lines; \citealt{Kollatschny1992}), and radio (\citealt{Ulvestad1999}).

\begin{figure*}[htp]
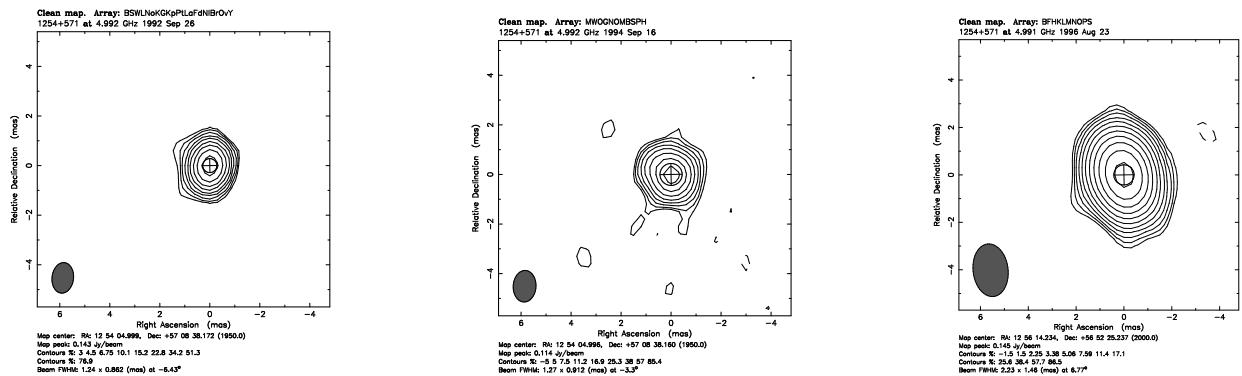

\begin{center}
  \includegraphics[width=0.23\textwidth,angle=0]{1254+571_1.eps}\hspace{50pt}
  \includegraphics[width=0.23\textwidth,angle=0]{1254+571_2.eps}\hspace{50pt}
  \includegraphics[width=0.22\textwidth,angle=0]{1254+571_3.eps}
  \caption{Three epochs of VLBI observations at 5 GHz of the object 1254+571, Mrk 231. Contour maps with maximum flux at 0.143 Jy/beam. Maps reproduced from \citet{Britzen2007b}.}
  \label{fig:Mrk231_radiomaps}
\end{center}
\end{figure*}

Mrk 231 was observed in three epochs (Sep 1992, Sep 1994, and Aug 1996) with the VLBA. The source remained unresolved during all three epochs (see Fig. \ref{fig:Mrk231_radiomaps}). \citet{Ulvestad1999} and \citet{Ulvestad1999b} however detected a pc-scale jet consisting of one component moving at sub-luminal speeds. \citet{Carilli1998} present strong arguments for the existence of a sub-kpc molecular disk formed in Mrk 231. The above properties point to a young AGN, created by a recent merger that has formed a molecular disk fueling gas into the nucleus of the newly formed AGN and a young radio jet, which agree well with the merger-driven evolution scenario (see also, \citealt{Hickox2009}).

We note that 1146+596 is somewhat similar to Mrk 231. This low redshift (z=0.011) radio galaxy appears to be extremely luminous in the near infrared (\citealt{Skrutskie2006}). It shows signs of ongoing interaction (\citealt{Pustilnik2001}) to which the infrared excess can be attributed. Moreover, \citet{Peck1998} detect redshifted H I gas in absorption, which appears to represent infall of gas towards the central engine.

In the above context, for the merger phase (in Table \ref{tab:evolution_candidates}) we select sources that appear disturbed and also show signs of ongoing starburst activity (Mrk 231, 3C 84, Mrk 501, 3C 371). These are primary candidates for BBH that have not yet sunk into the center of the merging system (similar to NGC 6240). Most of these sources exhibit variability at different wavelengths. Three sources (0316+413, 1418+546, 1652+398) show almost periodic variability in the radio. 0316+413 and 1652+398 are in addition members of clusters and therefore good candidates to be multiple merger systems.

We note here the ambiguity of the classification of sources into the different pre-merger stages (Table \ref{tab:evolution_candidates}). Starburst activity is the defining, although not unambiguous, property of identifying systems that are currently undergoing a merger, the alternative evidence discussed above being a weaker selection criteria.

\subsection{Binary black holes in the CJF}

We find 9 sources in the CJF sample that have been previously argued to be good BBH candidates. We present these sources in Tables \ref{tab:BBHkin} and \ref{tab:BBHrad}, along with the relevant references.

We compare the luminosities of these systems individually with the average luminosities at different wavelengths of a subsample of the CJF with similar median redshifts. All 9 sources are found to be significantly more luminous in X-rays. Two of these 9 sources (4C 37.11, BL Lac) are found to have above average K-band luminosity, possibly indicating a younger age for these sources. These two sources are fainter in X-rays than the remaining 7 sources. According to \citet{Sanders1988}, after a merger has taken place, the peak of the SED of the system is pushed towards higher frequencies, as the AGN blows away any obscuring material remaining from the merging phase. This is another indication that 4C 37.11 and BL Lac are indeed in an earlier evolutionary stage than the other BBH candidates.
\begin{table*}
\begin{center}
\caption{Kinematic and morphological information for CJF sources already considered to be BBH candidates. Columns (1)-(4) as in previous tables, Col. (5) shows the sources with X-ray detected jet, Col. (6) gives the misalignment angle between the pc and the kpc-scale jet of the source, Col. (7) indicates the bending as calculated in \citet{Britzen2008} for the inner-most component identified in the jet, Col. (8) gives references related to the environment of the source such as a companion or an interaction with other systems, Col. (9) comments on the general kinematics of the pc-scale jet, Col. (10) gives the reference for possible precession models for the source, and Col. (11) gives the references in regard to possible BBH. For references, see Table \ref{tab:BBHkinref}.}
\label{tab:BBHkin}
\begin{tabular}{l l l l l l l l l l l}
\hline\hline
\textbf{Source}	&	\textbf{Other Name}	&\textbf{Type}	    &\textbf{z}		&\textbf{X-ray Jet}	  &$\mathbf{\Delta}$\textbf{PA}	 &\textbf{Bending}	 &\textbf{Environment}	 &\textbf{Kinematics}		&\textbf{Precession}   		 &\textbf{BBH}	\\
                &                           &                   &               &                     &($\circ$)    &($\circ$)  &    &   &  & \\
\hline
0316+413&	3C 84		&G	    &0.018	&	          &20		 &51.9	&92		    &inward$^{101}$        &      &108	\\
0402+379&	4C 37.11	&G	    &0.055	&			  &27 / 12	 &		&2		    &stationary$^{102}$    &      &2,102\\
0716+714&			    &BL 	&0.300  &			  &75		 &2.6	&		    &stationary$^{103}$	   &106   &109	\\
1101+384&	Mrk 421		&BL 	&0.031	&			  &13		 &1.9	&82		    &slow$^{104}$          &      &110 	\\
1641+399&	3C345		&Q	    &0.595	&knot         &98		 &46.8	&88, 95		&                      &107   &111	\\
1652+398&	Mrk 501		&BL 	&0.034  &	          &83		 &10.5	&89, 99		&stationary$^{104}$    &106   &110 	\\
1803+784&			    &BL 	&0.680	&			  &70		 &18.0	&99		    &stationary$^{105}$    &      &112	\\
1928+738&	4C 73.18	&Q	  &0.302	&knot         &22		 &23.4	&	        &stationary$^{101}$    &106   &113	\\
2200+420&	BL Lac		&BL 	&0.069	&			  &30		 &24.8	&100		&                      &      &114	\\
\hline
\end{tabular}
\end{center}
\end{table*}

\begin{table*}
\begin{center}
\caption{References for Table \ref{tab:BBHkin}.}
\label{tab:BBHkinref}
\begin{tabular}{l l l l}
\hline\hline
\textbf{Symbol}   &    \textbf{Reference}              &\textbf{Symbol} &      \textbf{Reference}               \\
\hline
100      &    \citet{Disney1974}        &108    &   \citet{Pronik1988}      \\
101      &    \citet{Britzen2007b}      &109    &   \citet{Nesci2005}       \\
102      &    \citet{Maness2004}        &110    &   \citet{dePaolis2003}    \\
103      &    \citet{Britzen2006}       &111    &   \citet{Lobanov2005}     \\
104      &    \citet{Kellerman2003}     &112    &   \citet{Roland2008}      \\
105      &    \citet{Britzen2008}       &113    &   \citet{Roos1993}        \\
106      &    \citet{Lu2005}            &114    &   \citet{Zier2007}        \\
107      &    \citet{Caproni2004}       &       &                           \\
\hline
\end{tabular}
\end{center}
\end{table*}

\begin{table*}
\begin{center}
\caption{Variability information and, if available, timescales concerning CJF sources already considered to be BBH candidates, Col. (4) in the radio, Col. (5) in the infrared, Col. (6) in the optical, Col. (7) in the X-rays, and Col. (8) in the $\gamma$-rays. Columns (1)-(3) as before. For references, see previous tables.}
\label{tab:BBHrad}
\begin{tabular}{l l l l l l l l l l l l l l}
\hline\hline
\textbf{Source}	&	\textbf{Type}	&\textbf{z}		&\textbf{Var}$\mathbf{_{Radio}}$	&\textbf{Var}$\mathbf{_{IR}}$	 &\textbf{Var}$\mathbf{_{Opt}}$		  &\textbf{Var}$\mathbf{_{X}}$  	&\textbf{Var}$\mathbf{_{\gamma}}$ \\
	    &		&		    & (yr)		&		&  (yr)		           & (days)	& (days)	    \\
\hline
0316+413&	G	&0.018		&9.7		&		&			           &	  	&       	\\
0402+379&	G	&0.055		&yes		&		&			           &	  	&       	\\
0716+714&	BL	&0.3		&5.4-15		&yes	&3.3		           &	  	&yes    	\\
1101+384&	BL	&0.0308		&3.7-8.0 	&yes	&6.2		           &62	    &      	 	\\
1641+399&	Q	&0.595		&5.1, 3.5-4	&yes	&5.6, 11.4, 10.1, 1.9  &	    &      	 	\\
1652+398&	BL	&0.03366	&6.8-12.3	&yes	&yes               	   &23	    &23.2  	 	\\
1803+784&	BL 	&0.6797		&7.0-11.3	&		&0.003			           &	    &      	 	\\
1928+738&	Q	&0.302		&7.7-8.2	&		&yes		           &	    &      	 	\\
2200+420&	BL	&0.0686		&3.9-7.8	&yes	&7.8, 0.6, 0.88, 14, 7.7  &	    &      	 	\\
\hline
\end{tabular}
\end{center}
\end{table*}

In addition, we note that all BBH candidates exhibit periodicities across the spectrum. We find that 7 out of the total of 9 sources have bent VLBA jets. Out of the sources in Table \ref{tab:BBHkin}, 6 sources belong to a cluster or have a companion, illustrating the higher probability of these sources to harbor a BBH system.

\section{Conclusions}
\label{sec:conclusions}
The focus of this work has been twofold: (i) to investigate the evolutionary connection between active galaxies and merger events, and (ii) identify AGN candidates at different evolutionary stages, including binary black hole system candidates. To this end, we collect and present multiwavelength information about the CJF sources.

From the previous analysis, it becomes apparent that mergers or merger-induced effects (starburst activity) are ubiquitous in active galaxies. With almost 30\% of our sample showing evidence of interactions at z$<0.4$ and with indications that more such systems exist, although not yet observed, mergers appear to play a predominant role in the evolution not only of galaxies but AGN as well. We note here the importance of the absence of any biases towards transitory systems in the selection criteria of our sample. This supports a direct link between merger events and active galaxies in general. Furthermore, we show that multiwavelength variability may be linked to transitory systems. Assuming that multiwavelength variability originates in binary black hole systems, this provides an additional link between merger events, binary black holes, and active galaxies, as well as a possible identifier of these sources.

Following the merger-driven evolution scheme, we have also presented a first subsample of 28 CJF sources that are good tracers of evolution in AGN. We find at least 14 promising BBH candidates at different points in their evolution (5 in an unbound and 9 in a hard state). Classification of these sources cannot be unambiguous, given the lack of complete information. Additional observations are needed to clarify their true nature.

Following arguments from both analytical and numerical simulations, it is anticipated that binary black holes should form during merger processes. We show that at least 13.6\% of the CJF sources exhibit variability at different wavelengths (not including IDV), most of them being almost periodic. We show that most of the previous BBH candidates are highly variable across the spectrum, showing the tight connection between the two phenomena. The morphology and evolution properties of the AGN jets can also be linked to the evolution scheme and BBH systems. The morphology of CJF jets and its evolution will be studied in detail in a future paper.

More specifically, we find that:
\begin{itemize}
\item one CJF source (Mrk 231) has been classified as an ULIRG. Its properties fit well in the merging paradigm of AGN. We also note that the similar source 1146+596, which has a high luminosity in the near-infrared, is probably a merging system. Both sources are primary BBH candidates.
\item a correlation between radio luminosity and NIR luminosity exists. This is qualitatively explained in terms of reradiation of the non-thermal AGN emission. Low luminosity AGN appear to have a near-constant NIR luminosity.
\item 4.4\% of the sample exhibit distortions in their optical images, signs of ongoing or recent interaction, and that 3.1\% possibly have a companion. For $z<0.4$, we have found that 28.6\% of the CJF sources have either distortions or a companion, in excellent agreement with \citet{Hutchings1983} and \citet{Surace1998}.
\item the above listed conclusion, combined with the unbiased selection criteria of the CJF sample towards transitory systems, imply that merger events are indeed important to the evolution of active galaxies
\item multiwavelength variability appears to be linked to systems undergoing a transition from a thermal to non-thermal emission-dominated spectrum. This may be connected to the presence of a bound binary black hole in the nucleus of these systems.
\item following the categorization of \citet{Lotz2008}, we have selected the best candidate sources from our sample for different evolutionary phases. We have also selected sources as the best candidates in our sample for BBH systems.
\end{itemize}

More in-depth optical and infrared observations are needed, to unambiguously distinguish between different phenomena. High quality infrared data (mid and far) of our objects would provide us with the means to pinpoint the true position of each object on the assumed evolution track of galaxies. Alternatively, high resolution observations of those sources with multiwavelength variability would help us to model the possible existence of a binary black hole in these systems.

\begin{acknowledgements}
M. Karouzos was supported for this research through a stipend from the International Max Planck Research School (IMPRS) for Astronomy and Astrophysics. M.K. wants to thank Jens Zuther, Mar Mezcua, and Chin Shin Chang for insightful discussions. The authors also want to acknowledge the very helpful comments from an anonymous referee that substantially improved this paper. This research has made use of the NASA/IPAC Extragalactic Database (NED) which is operated by the Jet Propulsion Laboratory, California Institute of Technology, under contract with the National Aeronautics and Space Administration. This research has made use of NASA's Astrophysics Data System Bibliographic Services. This research has also made use of the VizieR service (\citealt{Ochsenbein2000}).
\end{acknowledgements}

\appendix
\section{Notes on individual sources}
\label{sec:individual}
Here we present additional information about individual sources from our subsample in Table \ref{tab:evolution_candidates}. For references, we refer to the respective tables.
\\*

\noindent\textit{\underline{Pre-merger:}}\\*
We classify as pre-merger sources systems that have a companion but do not have distorted morphologies:
\\*
\textbf{0954+658} (BL Lac, $z=0.368$): Besides its companion (at a projected separation 474 kpc), this source exhibits periodicity in the optical, and variability in $\gamma$-rays. It has a bent jet on both VLA and VLBA scales. This is a misaligned source. The nature of this object is ambiguous, as it may also be classified as being in the post-merger phase.\\*
\textbf{2021+614 (OW 637)} (quasar, $z=0.227$): Has a companion at a projected distance of 36 kpc. Its VLBA jet exhibits peculiar morphology, while its VLA jet is unresolved.\\*
\textbf{2352+495 (OZ 488)} (radio galaxy, $z=0.237$): Has a companion at a separation of 47 kpc. Its VLBA jet is bent, and VLA jet is unresolved.
\\*

\noindent\textit{\underline{First pass:}}\\*
We classify as first pass sources systems that have distorted morphologies but neither a visible companion nor signs of ongoing starburst activity:
\\*
\textbf{0018+729} (radio galaxy, $z=0.821$): This object appears to be distorted in the optical. On VLA scales, there is no jet resolved, while on VLBA scales the jet appears to be bent. \\*
\textbf{0108+388 (OC 314)} (radio galaxy, $z=0.669$): Source with disturbed morphology in the optical. Exhibits a straight jet on both VLBA and VLA scales.\\*
\textbf{0804+499 (OJ 508)} (quasar, $z=1.432$): Distorted morphology in the NIR. This source is variable in the radio. Exhibits bent VLBA jet and diffuse emission on VLA scales.\\*
\textbf{1146+596 (NGC 3894)} (LINER, $z=0.011$): Disturbed morphology in the optical. This source is positioned in the thermal regime on the NIR color-color diagram (see Fig. \ref{fig:color_diagram_candidates}), confirming its transitionary phase. The above indicates that the system has possibly undergone, or is undergoing, a merger. The radio jet appears straight. A VLA jet is unresolved.\\*

\noindent\textit{\underline{Max separation:}}\\*
In this phase, we classify sources that show distorted morphologies along with the existence of a companion, but no ongoing starburst activity. Distinguishing between pre-merger, first-pass, and max separation phases can be ambiguous:
\\*
\textbf{0402+379 (4C 37.11)} (radio galaxy, $z=0.055$): This object is both disturbed (in the optical) and has a close companion galaxy (projected distance 29 kpc). Its VLBI jet is bent. Diffuse radio emission is seen on VLA scales. Variable in the radio. It is positioned in the thermal regime of the color-color plot. This object can also be classified as a transitory object between the max separation and merger phase.\\*
\textbf{0710+439 (OI 417)} (Seyfert, $z=0.518$): Disturbed  optical morphology source with companion (projected distance 53 kpc). Bent VLBA jet.\\*
\textbf{0831+557 (4C 55.16)} (LINER, $z=0.24$): Possibly optically disturbed source with a companion (at a separation of 81 kpc). On VLBA scales, the jet appears bent, while on VLA scales the jet and counter-jet appear misaligned.\\*
\textbf{1031+567} (Seyfert 2, $z=0.46$): Disturbed morphology in the NIR. A companion was found in the optical (projected separation of 65 kpc). Shows straight VLBA jets. Unresolved VLA jet.\\*
\textbf{1823+568 (4C 56.27)} (BL Lac, $z=0.664$): Source exhibiting non-relaxed isophotes in the NIR, indirect sign of past merger activity. A possible close companion is also observed (projected distance of 46 kpc, no redshift information). The source is variable in the radio and was claimed to show almost periodicity. On VLBA scales, the jet appears fairly straight, while on VLA scales the jet appears to be misaligned with the inner jet.\\*
\textbf{1843+356} (radio galaxy, $z=0.764$): Disturbed morphology source (optical) with companion (projected distance 129 kpc). Straight VLBA jet, unresolved on VLA scales.\\*
\textbf{1946+708} (radio galaxy, $z=0.101$): This object appears to be disturbed in the optical. A possible companion galaxy (no redshift information available) is also observed at a projected distance of $\sim67$ kpc, thus we classify it as a max separation object. Its VLBA jet is highly bent. In the color-color diagram, it is positioned in the non-thermal regime. This may be caused by a misclassification. That the source is a merger remnant is the most plausible alternative.

\noindent\textit{\underline{Merger:}}\\*
In the merger phase, we classify sources that exhibit distorted morphologies and also ongoing starburst activity:
\\*
\textbf{0248+430} (quasar, $z=1.31$): This source exhibits signs of starburst activity (e.g., \citealt{Dudik2005}), while at the same time it is very luminous in both the optical and the X-rays. At a redshift $z=1.31$, the source is not optically resolved.\\*
\textbf{0316+413 (3C 84)} (radio galaxy, $z=0.018$): Disturbed morphology source (optical) with companion (projected distance 600 kpc). This source is positioned in the thermal region of the color-color plot and shows ongoing starburst activity. Both the VLBA and VLA jet of this source appear bent. This is a primary candidate for a merging system. It is periodically variable in the radio, optical, and X-rays.\\*
\textbf{1254+571 (Mrk 231)} (quasar, $z=0.042$): This object is an ULIRG and is positioned in the transitory region of the color-color diagram. It exhibits starburst activity (see Sect. \ref{sec:discussion} for more details). There is no jet resolved on either VLBA or VLA scales. Mrk 231 fits well into the evolutionary scenario proposed by \citet{Sanders1988} occupying a transitory phase between a merging system and an active galaxy. Combined with its variability (radio, optical, and X-rays), we argue that this is a late merger object, moving into the post-merger phase.\\*
\textbf{1652+398 (Mrk 501)} (BL Lac, $z=0.034$): This source is almost periodically variable at all wavelengths checked. It exhibits a disturbed morphology (optical) and is positioned in the thermal region of the color-color diagram. This system appears to be undergoing a series of mergers, supported by it being a member of a cluster. Its VLBI jet is bent, with diffuse emission on VLA scales.\\*
\textbf{1807+698 (3C 371)} (BL Lac, $z=0.051$): Optically disturbed source with a companion (projected separation 81 kpc). It is almost periodical in the optical. Variable in the infrared. The VLBA jet of this object appears straight. The VLA jet however appears bent and diffuse emission is present.

\noindent\textit{\underline{BBH / Post-merger candidates:}}\\*
Here we select sources that exhibit variability in at least three wavelength regimes (almost periodic in at least one wavelength). Correlated variabilities are one of the strongest pieces of evidence for a BBH. These systems may also be identified as post-merger systems, as we expect these systems to contain a hard BBH system in their nuclei.
\\*
\textbf{0219+428 (3C 66A)} (BL Lac, $z=0.444$): This object is positioned in the transitory region of the color-color diagram, thus is classified as an early post-merger system. Both the VLBI and VLA jet appear straight (with the VLBA jet exhibiting one sharp bend). It manifests almost periodic variability in the optical, and variability at radio and X-ray wavelengths.\\*
\textbf{0716+714} (BL Lac, unknown z): This source is variable in all wavelength regimes observed, with correlated variability on longer timescales in both the radio and the optical. It is a misaligned source with a bent jet on both VLBA and VLA scales.\\*
\textbf{0814+425 (OJ 425)} (BL Lac, $z=0.245$): Almost periodical variability in the radio and optical. Variable in the infrared. This system is positioned in the transitory region of the color-color diagram. Its VLBA jet appears bent.\\*
\textbf{0923+392 (4C 39.25)} (quasar,  $z=0.699$): This source exhibits variability at different wavelengths (almost periodicity in the radio). Its VLBI jet is bent. No jet is resolved on VLA scales. It is positioned in the non-thermal region of the color-color diagram.\\*
\textbf{1101+384 (Mrk 421)} (BL Lac, $z=0.031$): VLBA jet appears to be straight. The VLA jet appears bent. It is almost periodically variable in the radio, optical, and X-rays. It is variable in the infrared.\\*
\textbf{1418+546 (OQ 530)} (BL Lac, $z=0.151$): This object has a disturbed morphology. That it is positioned in the transitionary region of the color-color diagram and it exhibits periodicities in different wavelengths are both indicative of a post-merger galaxy. Its VLBI jet does not exhibit bends. It shows almost periodical variability in the radio. It is also variable in the infrared and optical.\\*
\textbf{1633+382 (4C 38.41)} (quasar, $z=1.807$): Misaligned source with bent jet on both VLBA and VLA scales. Almost periodic in radio. Variable in optical and $\gamma$-rays.\\*
\textbf{1641+399 (3C 345)} (quasar, $z=0.595$): This source appears to have a close companion (at a projected separation of 47 kpc) and is also member of a cluster, implying the possibility of multiple mergers. This source has already been modeled as BBH (\citealt{Lobanov2005}). The bent jet agrees well with the presence of a BBH system. Almost periodical in radio, optical, and X-rays. Variable also in the infrared.\\*
\textbf{2200+420 (BL Lac)} (BL Lac, $z=0.069$): This system is positioned in the non-thermal regime of the color-color diagram and appears almost periodical in radio and optical. It is also variable in infrared and $\gamma$-rays.

\bibliographystyle{aa}
\bibliography{bibtex}




\end{document}